\documentclass[a4paper,11pt]{article}
\pdfoutput=1 

\usepackage{jcappub} 

\usepackage[T1]{fontenc} 

\newcommand\aap{A\&A}                
\newcommand\apj{ApJ}                 
\newcommand\apjl{ApJ}                
\newcommand\apjs{ApJS}               
\newcommand\araa{ARA\&A}             
\newcommand\mnras{MNRAS}             
\newcommand\nar{New~Astron.~Rev.}    
\newcommand\nat{Nature}              
\newcommand\prd{Phys. Rev.~D}        
\newcommand\pasp{PASP}               
\newcommand\pasj{PASJ}               

\title{\boldmath Impact of radiative cooling on the magnetised geometrically thin 
accretion disc around Kerr black hole}

\author[a,1]{Indu K. Dihingia,\note{Corresponding author.}}
\author[a,b,c,d]{Yosuke Mizuno,}
\author[e,d,f]{Christian M. Fromm}
\author[g]{and Ziri Younsi}

\affiliation[a]{Tsung-Dao Lee Institute, Shanghai Jiao-Tong University, Shanghai, 520 Shengrong Road, 201210, People's Republic of China}
\affiliation[b]{School of Physics \& Astronomy, Shanghai Jiao-Tong University, Shanghai, 800 Dongchuan Road, 200240, People's Republic of China}
\affiliation[c]{Key Laboratory for Particle Physics, Astrophysics and Cosmology, Shanghai Key Laboratory for Particle Physics and Cosmology, Shanghai Jiao-Tong University,800 Dongchuan Road, Shanghai, 200240, People's Republic of China}
\affiliation[d]{Institut f\"{u}r Theoretische Physik, Goethe Universit\"{a}t, Max-von-Laue-Str. 1, 60438 Frankfurt am Main, Germany}
\affiliation[e]{Institut f\"ur Theoretische Physik und Astrophysik, Universit\"at W\"urzburg, Emil-Fischer-Str. 31, D-97074 W\"urzburg, Germany}
\affiliation[f]{Max-Planck-Institut f\"ur Radioastronomie, Auf dem H\"ugel 69, D-53121 Bonn, Germany}
\affiliation[g]{Mullard Space Science Laboratory, University College London, Holmbury St. Mary, Dorking, Surrey, RH5 6NT, UK}

\emailAdd{ikd4638@sjtu.edu.cn}
\emailAdd{mizuno@sjtu.edu.cn}
\emailAdd{christian.fromm@uni-wuerzburg.de}
\emailAdd{z.younsi@ucl.ac.uk}

\abstract{It is believed that the spectral state transitions of the outbursts in X-ray binaries (XRBs) are triggered by the rise of the mass accretion rate due to underlying disc instabilities. Recent observations found that characteristics of disc winds are probably connected with the different spectral states, but the theoretical underpinnings of it are highly ambiguous. To understand the correlation between disc winds and the dynamics of the accretion flow, we have performed General Relativistic Magneto-hydrodynamic (GRMHD) simulations of an axisymmetric thin accretion disc with different accretion rates and magnetic field strengths.
Our simulations have shown that the dynamics and the temperature properties depend on both accretion rates and magnetic field strengths. We later found that these properties greatly influence spectral properties. We calculated the average coronal temperature for different simulation models, which is correlated with high-energy Compton emission. Our simulation models reveal that the average coronal temperature is anti-correlated with the accretion rates, which are correlated with the magnetic field strengths. We also found that the structured component of the disc winds (Blandford-Payne disc wind) predominates as the accretion rates and magnetic field strengths increase. In contrast, the turbulent component of the disc winds ($B_{\rm tor}$ disc wind) predominates as the accretion rates and magnetic field strengths decrease. Our results suggest that the disc winds during an outburst in XRBs can only be understood if the magnetic field contribution varies over time (e.g., MAXI J1820+070).}

\begin{document}
\maketitle
\flushbottom

\section{Introduction}

Black hole X-ray binaries (BH-XRBs) are one of the most observed and studied astrophysical sources in the literature. They are primarily observed during their outbursts. In an outburst, the source undergoes a cycle of processes known as the `Q' diagram \citep{Belloni-etal2011, Belloni-Motta2016, Ingram-Motta2019}. Understanding different parts of the `Q' diagram (hard, intermediate, and soft states) is becoming less and less ambiguous with time, although the transitions between different parts still need to be explored in great detail.  

An outburst can be classified into two phases, i.e., the rising and decaying phases. A hard state is observed in both phases. The common consensus is that the accretion rate alone triggers the spectral state transitions in X-ray binaries \citep[see e.g.,][for review]{Narayan-etal1998, Yuan-Narayan2014}. However, the driving mechanism of the accretion is likely to be magnetic in nature \citep[e.g.,][]{Balbus-Hawley1998, Dihingia-etal2023b}. Very recent observations try to couple outflows/wind with the different spectral states of X-ray binaries \citep[e.g.,][]{Gallo-etal2005,Miller-etal2006,Ponti-etal2012,Neilsen-etal2013, DazTrigo-Boirin2016,Tetarenko-etal2018}. In the black hole X-ray binaries MAXI~J1820+070 and MAXI~J1803-298, many observations reported clear signatures of disc-wind in both the hard states in its 2018 outburst \citep{Munoz-Darias-etal2019, Mata-Sanchez-etal2022}. However, the rising hard state of MAXI~J1820+070 shows a higher degree of optical polarisation than the decaying hard state, which is found in an intrinsically unpolarised state \citep{Kosenkov-etal2020}. Possibly, the rising phase is dominated by Blandford-Payne mechanism-driven disc-wind (BP disc-wind, structured magnetic field)\citep{Blandford-Payne1982}, whereas in the decaying phase, turbulent toroidal magnetic field-driven disc-wind or thermal pressure-driven disc-wind may dominate (unstructured magnetic field). Keeping all of these in mind, it is possible that XRB's disc winds are the missing piece to the puzzle for figuring out how the spectral state transition happens during an outburst. These facts hint that during an outburst, the strength and structure of the magnetic field are expected to change along the timeline. Therefore, in general, both the accretion rate and the magnetic field configuration may be important for the understanding of an outburst of XRBs.

It is clear from the above discussion that the underlying disc winds and their properties may help infer the triggers for the spectral state transitions in X-ray binaries. Earlier thin-disc general relativistic magnetohydrodynamic (GRMHD) simulations show different components of disc-winds from the accretion disc \citep{Vourellis-etal2019, Vourellis-Fendt2021}. Later, \cite{Dihingia-etal2021, Dihingia-Vaidya2022} found strong signatures of BP-dominated disc winds with strong and inclined magnetic field configurations. \cite{Dihingia-etal2022b} also suggested that a truncated accretion disc may be suitable for understanding the rising phase of the `Q' diagram. However, in these studies, they do not consider any contribution from the radiative cooling processes, which is very consequential in the context of outbursts. Radiative cooling becomes particularly unavoidable during the outburst's soft and soft-intermediate states. Recently, \cite{Dexter-etal2021} showed that strongly magnetised accretion flow could produce the luminosity of a hard state with the sub-Eddington accretion rate. In a similar context, \cite{Wielgus-etal2022} have recently demonstrated using radiative GRMHD that the spectra of a geometrically thick disc (puffy disc) resemble the intermediate states between soft and hard emission states of BH-XRBs. In this paper, in order to understand the dynamical and spectral features of the accretion disc, we study thin accretion discs with radiative cooling with different magnetic field strengths and accretion rate limits. Finally, we try correlating them with a different part of the `Q' diagram.

This work follows \cite{Dihingia-etal2023} in terms of GRMHD with radiative cooling, where we consider a two-temperature accretion flow. The electrons in the fluids are subjected to radiative cooling by Bremsstrahlung, synchrotron, synchrotron self-comptonisation, and black body radiation. The electrons heat up due to the Coulomb collision and magnetic heating processes (turbulent and reconnection heating). This approach to studying accretion flow is not new for AGNs. The effects of radiative cooling on the accretion flow have been studied for several years, particularly for Sgr A$^*$ and M\,87 \citep{Sadowski-etal2017,Ryan-etal2017,Chael-etal2018,Ryan-etal2018,Chael-etal2019}. We perform the same for a thin-disc scenario, which is primarily applicable to BH-XRBs. In the next section (Section 2), we discuss the mathematical details of disc setup, radiative processes, etc. In Sections 3-7, we discuss results obtained from different simulation models and their interpretations and analyses. Finally, in Section 8, we conclude our results with a discussion on the astrophysical implications of our results and future plans.

Note that throughout the manuscript, we solve the equations and express the quantities in geometrised units, where $G=M_\bullet=c=1$. $G, M_\bullet$ and $c$ refer to the universal gravitational constant, the mass of the black hole, and the speed of light, respectively. In this system of units, mass, length, and time scales are presented in terms of $M_\bullet, r_g=GM_\bullet/c^2$, and $t_g=GM_\bullet/c^3$, respectively. 

\section{Model setup}

In this section, we discuss the mathematical formalism of this study. We have employed the GRMHD \texttt{code BHAC} \citep{Porth-etal2017,Olivares-etal2019} with an adaptive-mesh refinement (AMR) scheme to carry out 2D (axisymmetric) simulations of magnetised thin-discs around Kerr black holes. The governing equations of ideal GRMHD in terms of conservation laws are as follows \citep[see][]{DelZanna-etal2007, Rezzolla-Zanotti2013},
\begin{align}
    \partial_t\left(\sqrt{\gamma}\boldsymbol{U}\right) + \partial_i\left(\sqrt{\gamma}\boldsymbol{F^i}\right) = \sqrt{\gamma}\boldsymbol{S}\,.
    \label{eq:01}
\end{align}
In the presence of radiative cooling, the source term is modified as 
$\boldsymbol{ S}=\boldsymbol{S_0} + \boldsymbol{S^\prime}$, where $\boldsymbol{S^\prime}$ 
is the contribution of all the cooling and heating processes. The explicit form of the additional source term 
can be written as \citep{Dihingia-etal2023}, 
\begin{align}
    \boldsymbol{S^\prime}=
    \begin{pmatrix}
0 \\
-\alpha \gamma v_j \Lambda \\
-\alpha \gamma \Lambda \\
0 \\
\end{pmatrix}\,,
\label{eq:02}
\end{align}
where the symbols $v_j$, and $\Lambda$ denote the covariant fluid three-velocity, and the total radiation-cooling term, respectively. The other symbols have their usual meaning in ideal GRMHD, i.e., conserved variables $\boldsymbol{U}$, fluxes $\boldsymbol{F^i},$ and sources $\boldsymbol{S_0}$. The detailed expression of these variables in the case of ideal GRMHD without radiative cooling has been reported by \cite{Porth-etal2017} [see Eqs.~(23) and (30) there \citep[for detail follow][]{Rezzolla-Zanotti2013}]. Finally, $\alpha$ and $\gamma$ stand for the lapse function and the determinant of the induced three metrics, respectively. The explicit form of them in Kerr black hole is given by,
\begin{align}
    \alpha^2=\frac{\Delta \left(a^2 \cos ^2\theta+r^2\right)}{\left(a^2+r^2\right){}^2-a^2 \Delta \sin ^2\theta  },~~{\rm and}~~ \gamma=-g/\alpha^2,
\end{align}
where $\Delta=\left(a^2+r^2-2 r\right)$ and determinant of the metric $g=-\left(a^2 \cos ^2\theta +r^2\right)^2 \sin ^2\theta$. Equation~\ref{eq:01} describes the single-fluid MHD equations. To expand our investigation to a two-temperature 
framework, we additionally solve the electron-entropy equation including dissipative heating, Coulomb interactions, and 
radiation-cooling processes \citep[e.g.,][]{Sadowski-etal2017}, i.e., 
\begin{align}
    T_e \nabla_\mu \left(\rho u^\mu \kappa_e \right) = f_e^+ Q + \Lambda_{\rm ei} - \Lambda\,,
\label{eq:03}
\end{align}
where $\kappa_e := \exp[(\tilde{\Gamma}_e -1)s_e]$, and subsequently $s_e:=p/\rho^{\tilde{\Gamma}_e}$ is the electron entropy per particle. Here, $\tilde{\Gamma}_e$ represents the adiabatic index for the electrons. $f_e^+$ is the fraction of dissipating heating $(Q)$ transferred to the electrons, and $\Lambda_{\rm ei}$ is the rate of energy transferred to the electrons from the ions due to the Coulomb interaction. For current work, we consider a turbulent-heating prescription is used to determine the fraction of dissipative heating $f_e^+$ following \cite{Howes2010,Howes2011}. The explicit expression of $f_e^+$ is given as
\begin{align}
    f_e^+ := \frac{1}{1+Q_i/Q_e}\,,
\end{align}
where
\begin{align*}
    \frac{Q_i}{Q_e} = c_1 \frac{c_2 + \beta^\alpha}{c_2 + \beta^\alpha}
    \sqrt{\frac{m_p T_i}{m_eT_e}}\exp(-1/\beta)\,,
\end{align*}
where $m_e$ and $m_p$ referring to the mass of electron and proton, respectively, and subsequently, $c_1=0.92, c_2=1.6/(T_i/T_e), c_3 = 18 + 5 \log(T_i/T_e)$, $\alpha=2-0.2\log(T_i/T_e)$, and $\beta=p_{\rm
gas}/p_{\rm mag}$. In order to solve the electron entropy equation (Eq.~\ref{eq:03}), we assume charge neutrality, i.e., $n=n_e=n_i$ (equal number densities for electrons and ions), and also assume that the four-velocity (bulk velocity) of the electrons, ions, and the flow be the same, i.e., $u^\mu=u^\mu_e=u^\mu_i$. This does not have to be strictly true, but it is a plausible first approximation under the magnetohydrodynamic limit, where bulk motions are dominated by electromagnetic and gravitational forces. In such a highly magnetized and relativistic environment, the collective dynamics of the plasma ensure that both ions and electrons co-move, with negligible relative drift. While temperature differences could theoretically lead to some deviations, these effects are beyond the scope of GRMHD, which focuses on large-scale, collective behavior. 
For instance, in the solar wind, the relative speeds of different particle species can become close to the Alfven speed \citep{Bourouaine-etal2013}. However, a similar approach has been widely used for two-temperature frameworks in semi-analytical and numerical studies \citep[e.g.,][]{Narayan-Yi1995,Nakamura-etal1996,Manmoto-etal1997,Ressler-etal2015, Chael-etal2018, Dihingia-etal2018,Dihingia-etal2020}. 
In this work, we follow \cite{Ressler-etal2015} to treat dissipative-heating terms and update the electron's entropy explicitly following conservation principles. We solve all flow variables using Eqs.~\eqref{eq:01} and \eqref{eq:03} following \cite{Dihingia-etal2023}. By obtaining the internal energy of flow and electrons, we calculate the internal energy contribution of protons and subsequently calculate the temperatures of electrons ($T_e$) and ions ($T_i$). For simplicity, we consider the ideal equation of state with three adiabatic indexes for flow (gas as a whole), electrons, and ions as $\tilde{\Gamma}_g=13/9$, $\tilde{\Gamma}_e=4/3$, and  $\tilde{\Gamma}_p=5/3$ \citep[e.g.,][]{Shiokawa-etal2012,Ryan-etal2017,Ryan-etal2018,Dihingia-etal2023}. These choices are also supported by simulations with variable adiabatic indexes for each component \citep{Sadowski-etal2017}.

\subsection{Dissipative and radiative processes}
We consider that the electrons are subjected to radiative cooling and also consider Coulomb interaction between electrons and ions following \cite{Dihingia-etal2023}. We consider Bremsstrahlung ($Q_{\rm br}$) and synchrotron ($Q_{\rm cs}$) radiative cooling processes to be dominated in the optically thin regime. With this, the radiative cooling rate for the optically thin flow is calculated as:
\begin{align}
    Q^-=Q_{\rm br} + \eta \exp(-(\sigma/\sigma_{\rm cut})^2)Q_{\rm cs}\,,
    \label{eq:thincool}
\end{align}
where the Compton enhancement factor $(\eta)$, which is calculated following \citep{Narayan-Yi1995}. To prevent redundancy, we do not supply a detailed description of the cooling rates. See \cite{Esin-etal1996, Dihingia-etal2023} for more detailed expressions. To ensure that the highly magnetized region contributes negligibly to thermal synchrotron radiation, we multiply with the factor $\exp(-(\sigma/\sigma_{\rm cut})^2)$ to the synchrotron cooling rate.
The value of $\sigma_{\rm cut}$ may impact the calculated electron temperature in the funnel region ($\sigma>1$). If the emission from the funnel region is neglected, the choice $\sigma_{\rm cut}=10$ will not impact any dynamics and radiative properties outside the funnel region (since for $\sigma\le1$, $\exp(-(\sigma/\sigma_{\rm cut})^2)Q_{\rm cs}\sim Q_{\rm cs}$).
The equation \ref{eq:thincool} gives the total cooling due to optically thin matter. However, in the thin disc, 
the optically thick components can not be neglected due to the high optical depth. Therefore, we consider 
generalised cooling formula suggested by \cite{Narayan-Yi1995} and \cite{Esin-etal1996},
\begin{align}
    Q^{-}_{\rm tot} = \frac{4\sigma_T T_e^4/H}{1.5\tau + \sqrt{3} + \tau_{\rm abs}^{-1}}\,,
    \label{eq:totq}
\end{align}
where $\tau = \tau_{\rm es} + \tau_{\rm abs}$, and $\tau_{\rm abs}=(HQ^-/\sigma_T T_e^4)$. 
Finally, we included the Coulomb interaction and radiation-cooling terms in the governing equations in 
code units as $\Lambda_{\rm ei}=Q_{\rm ei}/U_{\rm c}$ and $\Lambda=Q^-/U_{\rm c}$, where 
$U_{\rm c}=\dot{M}_{\rm cgs}c^2/r_g^3$ and $\dot{M}_{\rm cgs}$ is the mass-accretion rate $(\dot{M})$ in CGS units. 
\subsection{Initial conditions}
For this work, we set up a geometrically thin accretion disc as the initial condition for the simulations following \cite{Dihingia-etal2021,Dihingia-Vaidya2022}, which is based on the standard thin-disc model proposed by \cite{Novikov-Thorne1973}. In this setup, the initial density distribution on the poloidal plane in Boyer-Lindquist (BL) coordinates is given by,
\begin{align}
\rho(r,\theta) = \rho_e(r) \exp\left(-\frac{\alpha^2 z^2}{H_e^2}\right); ~~ z=r\cos(\theta)\,.
\label{eq-rho}
\end{align}
To maintain the geometrical thin nature of the initial disc, we choose $\alpha=2$ and equilibrium disc-height $(H_e)$ following \cite{Riffert-Herold1995} and \cite{Peitz-Appl1997}. In Eq.~\eqref{eq-rho}, $\rho_e(r)$ provides the density profile on the equatorial plane, which is given by,
\begin{align}
\rho_e(x)=\left(\frac{\Theta_0}{\cal K}\right)^{1/(\tilde{\Gamma}_g -1)} 
              f(x)^{1/(4(\tilde{\Gamma}_g - 1))}\,,
\label{eq-rhoe} 
\end{align}
where $x=\sqrt{r}$, $\Theta_0$ is a constant that fixes the initial temperature distribution of the thin disc. Here, 
we chose $\Theta_0 = 0.001$. For this study, we consider the entropy constant ${\cal K}=0.1$. Finally, $f(x)$ is calculated as,
\begin{align}
\begin{aligned}
f(x) =& \frac{3}{2x^2} \frac{1}{ x^2 \left(2 a+x^3-3 x\right)}\bigg[ x - x_0 -\frac{3}{2}\ln\left(\frac{x}{x_0}\right) \\
&- \frac{3\left(l_1-a\right)^2}{l_1(l_1-l_2)(l_1-l_3)} \ln \left(\frac{x- l_1}{x_0-l_1}\right) \\
&- \frac{3\left(l_2-a\right)^2}{l_2(l_2-l_1)(l_2-l_3)}\ln \left(\frac{x- l_2}{x_0-l_2}\right) \\
&- \frac{3\left(l_3-a\right)^2}{l_3(l_3-l_1)(l_3-l_2)}\ln \left(\frac{x- l_3}{x_0-l_3}\right)\bigg]\,, \\
\end{aligned}
\label{eq-fx}
\end{align}
where $x_0=\sqrt{r_0}$, $r_0$ is the radius of the innermost stable circular orbit (ISCO), and $l_1, l_2,$ and $l_3$ are the roots of the cubic equation $l^3 - 3l + 2a=0$, the explicit forms of the roots can be found in \cite[][see Eq. 14]{Page-Thorne1974}. To completely describe the initial condition, we also need to supply the initial azimuthal velocity along with the density distribution, which is given as follows \citep{Dihingia-etal2021},
\begin{align}
u^\phi(r,\theta) = \left(\frac{\cal A}{{\cal B}+ 2 {\cal C}^{1/2}}\right)^{1/2}\,,
\label{eq-12}
\end{align}
where
$$
\begin{aligned}
{\cal A}=&\left(\Gamma^r_{tt}\right)^2,\\
{\cal B}=&g_{tt}\left(\Gamma^r_{tt}\Gamma^r_{\phi \phi}-2 {\Gamma^r_{t\phi}}^2\right)+2 g_{t\phi} \Gamma^r_{tt} \Gamma^r_{t\phi} - g_{\phi \phi } {\Gamma^r_{tt}}^2,\\
{\cal C}=&\left({\Gamma^r_{t\phi}}^2 - \Gamma^r_{tt} \Gamma^r_{\phi \phi}\right) (g_{t\phi} \Gamma^r_{tt}- g_{tt} \Gamma^r_{t\phi})^2.\\
\end{aligned}
$$
Here, $\Gamma^\alpha_{\beta\gamma}$ and $g_{\mu\nu}$ are the non-zero components of the Christoffel symbols and the metric around the Kerr black hole, respectively. Note that we use Modified Kerr-Schild (MKS) coordinates to solve the GRMHD equations. Accordingly, the initial conditions are transformed from BL coordinates to MKS coordinates correctly before being supplied to the simulation models. Explicit transformation equations can be found in \cite{McKinney-Gammie2004, Porth-etal2017}.

\subsection{Parametric Models}
Depending on our motivation, we devised a few simulation models for magnetised accretion flow around a Kerr black hole. We supply an initial large-scale poloidal magnetic field threading the accretion disc using a vector potential. The explicit expression of the adopted vector potential ${\cal A}_\phi$ on the poloidal plane is given by \cite{Zanni-etal2007},
\begin{align}
{\cal A}_\phi \propto \left(r \sin \theta\right)^{3/4} \frac{m^{5/4}}{\left(m^2 + \tan^{-2}(\theta-\pi/2)\right)^{5/8}}\,,
\label{eq-14}
\end{align}
where $m$ is a parameter determining the inclination of the initial magnetic field lines. The parameter $m$ is crucial in launching disc-wind from the accretion disc \citep{Blandford-Payne1982,Dihingia-etal2021}. For this study, we fixed the value of the inclination parameter to $m=0.1$. The magnetic field strength is set by supplying the input plasma-$\beta$ value, i.e., $\beta_0 = p_{\rm gas}^{\rm max}/p_{\rm mag}^{\rm max}$, where $p_{\rm gas}^{\rm max}$ and $p_{\rm mag}^{\rm max}$ are the maximum values of the gas pressure and the magnetic pressure, respectively. For a better understanding of the initial disc properties, we also calculate the maximum value of the plasma-$\beta$ parameter at the accretion disc ($\beta_{\rm max}$). Models \texttt{A}, \texttt{B}, \texttt{C}, and \texttt{D} are devised to understand the role of the input accretion rate ($\dot{m}=\dot{M}/\dot{M}_{\rm Edd}$) on the dynamics of the accretion flow. The values of the input accretion rates are given in the table \ref{tab-01} in terms of Eddington units (i.e., $\dot{M}_{\rm Edd} = 1.44\times10^{18}M_{\bullet}/M_{\odot}$\,gs$^{-1}$. The models \texttt{C}, \texttt{E}, \texttt{F}, and \texttt{G} are devised to understand the role of the magnetic field strength (supplying input $\beta_0$) on the dynamics of the thin accretion disc. Additionally, we included model \texttt{C-R} with the reconnection heating model \citep{Rowan-etal2017} to have qualitative comparisons with the turbulent heating model (model \texttt{C}) of the same accretion rate and magnetic field strength. The explicit values of the $\beta_0$ and $\beta_{\rm max}$ for these models are displayed in table~\ref{tab-01}. Last but not least, targeting accretion flow around a rotating stellar mass black hole, we fixed the mass and spin of the black hole as $M_{\bullet}=10M_{\odot}$ and $a=0.9375$, respectively.
\begin{table}
\centering
  \begin{tabular}{| l | c |c | c |}
    \hline
    Model & $\dot{m}$ & $\beta_0$ & $\sim\beta_{\rm max}$\\ 
    \hline
    \texttt{A} &  $0.1$  & 0.1 & $2\times10^3$ \\
    \texttt{B} &  $0.05$  & 0.1 & $2\times10^3$ \\
    \texttt{C} &  $0.01$  & 0.1 &  $2\times10^3$ \\
    \texttt{C-R} &  $0.01$  & 0.1 &  $2\times10^3$ \\
    \texttt{D} &  $0.001$  & 0.1 &  $2\times10^3$ \\
    \texttt{E} &  $0.01$  & 0.5 &  $1\times10^4$ \\
    \texttt{F} &  $0.01$ & 0.05 & $1\times10^3$ \\
    \texttt{G} &  $0.01$ & 0.01 & $2\times10^2$ \\
    \hline
  \end{tabular}
\caption{The explicit values of normalised accretion rates $\dot{m}$, input plasma-$\beta$, 
and $\beta_{\rm max}$ for different simulation models.}
\label{tab-01}
\end{table}

We solve the GRMHD equations on the poloidal plane $(r,\theta)$ in an axisymmetric consideration, where $r$ ranges from $r=r_{\rm H}$ to $r=500~r_g$ and $\theta$ ranges from $\theta=0$ to $\theta=\pi$. To do this, we adopt Modified Kerr-Schild (MKS) coordinates and divide our numerical domain with an effective resolution of $1024\times512$ (with base resolution $512\times256$). To make the simulation faster, we choose a static mesh refinement (SMR) grid with the highest level of resolution close to the black hole $r<50r_g$. The efficacy of the choice of resolution is discussed in Appendix A in terms of the MRI quality factor. For the purpose of better explanation and comparison, we consider model \texttt{C} to be the reference model. 

\section{Properties of inflow}
In this section, we discuss the timing and dynamic properties of the accretion flow close to the central black hole. Subsequently, we discuss their relationships with the accretion rates and magnetic field strengths.
\subsection{Temporal evolution}
\begin{figure}
    \centering
    \includegraphics[scale=0.32]{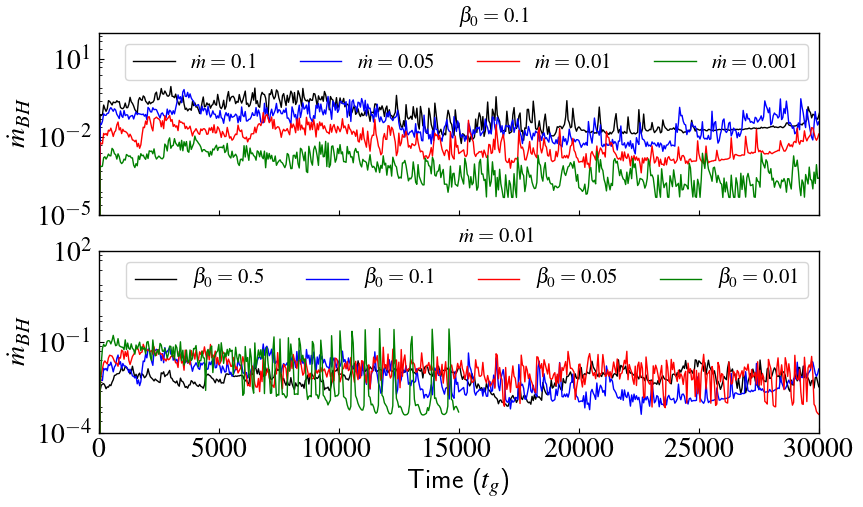}
    \includegraphics[scale=0.32]{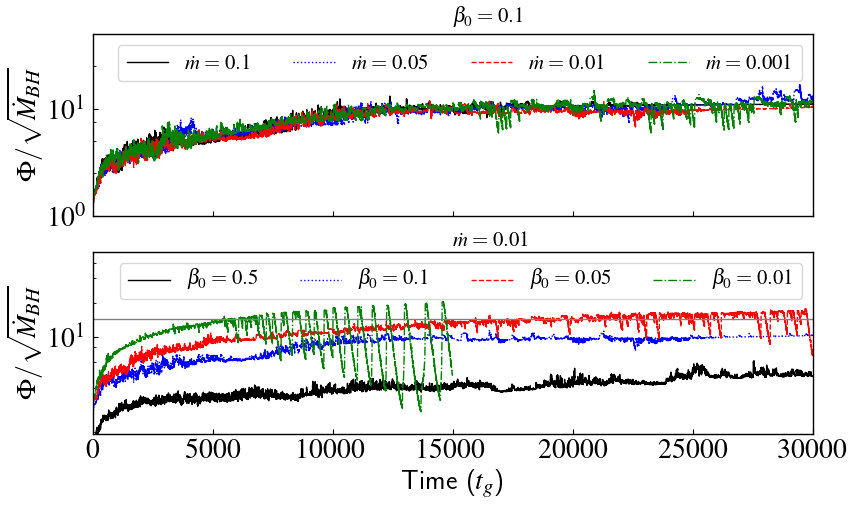}
    \caption{Plot of accretion rate measured at the event horizon$(\dot{M}_{\rm BH})$ profiles (left) and 
    normalized magnetic flux $(\Phi/\sqrt{\dot{M}_{\rm BH}})$ around the event horizon profiles (right) as a function of simulation time (in units of $t_g$)
    for different simulation models. The upper panels show accretion rate profiles by varying input accretion rates, 
    while lower panels show accretion rate profiles by varying magnetic field strength ($\beta_0$). 
    The thin horizontal line in the left figure corresponds to $\Phi/\sqrt{\dot{M}_{\rm BH}}=15$.
    }
    \label{fig:acc-mag}
\end{figure}

In this section, we study the long-term temporal behaviour of our simulation models. In Fig. \ref{fig:acc-mag}, we plot the accretion rate profiles measured at the event horizon in Eddington units $(\dot{m}_{\rm BH}=\dot{M}_{\rm BH}/\dot{M}_{\rm Edd})$ (left) and normalised magnetic flux $(\Phi/\sqrt{\dot{M}_{\rm BH}})$ profiles around the event horizon (right) for different simulation models, where $\dot{M}_{\rm BH}$ refers to the accretion rate calculated at the event horizon of the black hole. We follow Eqs.~(95) and (96) of \cite{Porth-etal2017} to calculate the $\dot{M}$ and $\Phi$. In the upper panels, we show variation with accretion rates with a fixed value of input plasma-$\beta$ parameter $(\beta_0=0.1$, marked on the top of the figure), while the lower panels show variations with magnetic field strength ($\beta_0$) for a fixed accretion rate ($\dot{m}=0.01$, marked on the top of the figure). In the lower-left panel, we mark $\Phi/\sqrt{\dot{M}_{\rm BH}}=15$ with a horizontal line, which denotes the maximum saturation flux for a magnetically arrested disc (MAD) configuration in the code units \citep[e.g.,][]{Porth-etal2021,Mizuno-etal2021}. Note that this value differs by a factor $\sqrt{4\pi}$ from the commonly used definition \citep{Tchekhovskoy-etal2011,Tchekhovskoy-etal2012,McKinney-etal2012} due to the choice of the simulation framework. We observe that the accretion rate profile in all our simulation models shows a quasi-steady behaviour with time, except for model \texttt{G} ($\beta_0=0.01, \dot{m}=0.01)$. In model \texttt{G}, the accretion rate profile initially reaches a quasi-steady state, but after a simulation time $t>6000\,t_g$, the accretion rate profile shows sporadic oscillations. These oscillations are similar to the ones observed by \cite{Dihingia-etal2021}.

With the increase in the input accretion rate ($\dot{m}$), the accretion rate profiles show quite a similar trend. Although due to different normalisation factors, the average value in Eddington units increases with the input accretion rate parameter $(\dot{m})$. The normalised magnetic flux around the horizon $\Phi/\sqrt{\dot{M}_{\rm BH}}$ profiles is insensitive to the input accretion rate ($\dot{m}$). All the models in panel (b) show a monotonically increasing behaviour with time, and after $t\gtrsim10000\,t_g$, the profiles saturate with $\Phi/\sqrt{\dot{M}_{\rm BH}}\sim8$. 

With the increase in magnetic field strength (decreasing of $\beta_0$), the quasi-steady value of the accretion rate increases. Strong magnetic fields can facilitate stronger disc winds and consequently help in angular momentum transport, which results in a higher rate of mass flux to the black hole \citep{Dihingia-etal2021}. With the decrease of input $\beta_0$, the $\Phi/\sqrt{\dot{M}_{\rm BH}}$ shows scaling behaviour, except for $\beta_0=0.01$ (model \texttt{G}). For all the other models, the value of $\Phi/\sqrt{\dot{M}_{\rm BH}}$ saturates after simulation time $t\gtrsim10000\,t_g$. The saturation values for $\beta_0=0.5, 0.1,$ and  $0.05$ is $\Phi/\sqrt{\dot{M}_{\rm BH}}\sim2.5, 8$, and $15$, respectively. However, for model \texttt{G} ($\beta_0=0.01$, green line), the value of normalised magnetic flux exceeds $\Phi/\sqrt{\dot{M}_{\rm BH}}=15$ (see the horizontal line in panel d) after simulation time $t\gtrsim4000\,t_g$. After that, we observe large oscillatory behaviour in the magnetic flux profile, which signifies a magnetically arrested disc (MAD, \citep{Tchekhovskoy-etal2011}) configuration. During MAD, we also observe similar oscillations in the accretion rate profile for the same simulation model (see green line in panel c). However, it is important to note that MAD accretion flows in full three-dimensional (3D) studies show non-axisymmetric flow properties, which makes MAD more interesting \citep[e.g.,][]{Porth-etal2021,Ripperda-etal2022}. Accordingly, an axisymmetric framework cannot explain all the MAD-like behaviour of the flow. Following this, to avoid unreliable axisymmetric artefacts of MAD flow, we do not run the MAD models up to $t=30000\,t_g$, unlike all the other models. Thus, in our study, we only consider one such model, and all other simulation models do not develop MAD configurations; they are known as standard and normal evolution (SANE) configurations. In the future, while doing 3D simulations, we will explore MAD models in detail. In this study, the MAD model is considered only for reference and to hint at the possible results in dynamically important magnetic field cases.

\begin{figure}
    \centering
    \includegraphics[scale=0.35]{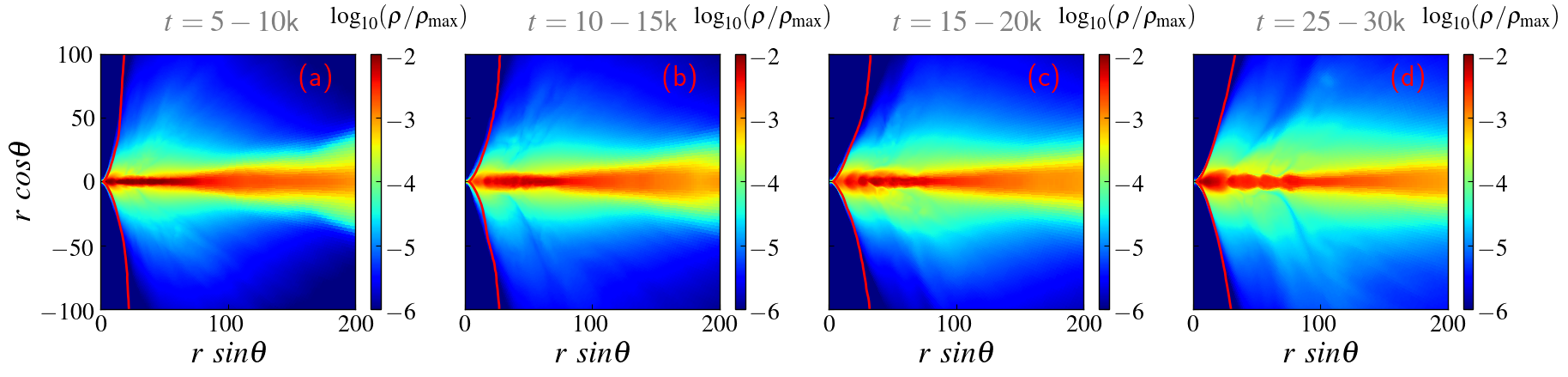}
    \caption{Distribution of time-averaged logarithmic normalised density distribution 
    $(\log_{10}(\rho/\rho_{\rm max}))$ for the reference model (model C) on the poloidal plane for 
    different simulation times, marked on the top of the figure. The solid red line corresponds to $\sigma=1$ contour.}
    \label{fig:density-time}
\end{figure}
\subsection{Density distribution}
\begin{figure}
    \centering
    \includegraphics[scale=0.6]{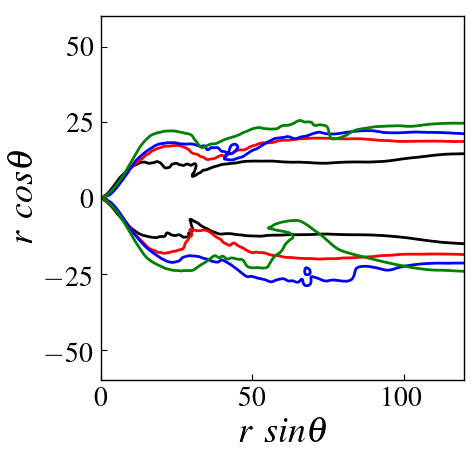}
    \caption{Plot of time-averaged high-density contours $(\rho/\rho_{\rm max}=0.01)$ for the reference 
    model at time $t=5000-10000\,t_g$ (black), $10000-15000\,t_g$ (red), 
    $15000-20000\,t_g$ (blue), and $25000-30000\,t_g$ (green), respectively.}
    \label{fig:density_con_time}
\end{figure}

In Fig.~\ref{fig:density-time}, we show the time-averaged logarithmic normalised density distribution of $(\log_{10}(\rho/\rho_{\rm max}))$ for different simulation times for the reference model (model \texttt{C}), where we consider the input accretion rate $\dot{m}=0.01$ and input plasma-$\beta$ parameter $\beta_0=0.1$. The red solid contour in the panels represents the boundary of $\sigma=b^2/\rho=1$. Panels (a), (b), (c), and (d) correspond to the distribution calculated between simulation times $t=5000-10000\,t_g$, $10000-15000\,t_g$, $15000-20000\,t_g$, and $25000-30000\,t_g$, respectively. The quasi-steady structure of the accretion disc can be divided into three major parts: (i) high-density thin-disc around the equatorial plane; (ii) low-density off-equatorial region; and (iii) very low-density funnel region. The low-density region contributes to the disc winds, and the very low-density region around the polar axis contributes to the relativistic Poynting-dominated jet \citep[see][]{Vourellis-etal2019,Dihingia-etal2021}. In order to understand the time evolution of disc structure better, we plot the contours of the edge of high-density region ($\rho/\rho_{\rm max} = 0.01$) at different time range in Fig.~\ref{fig:density_con_time}, where black, red, blue, and green lines correspond to $\rho/\rho_{\rm max}=0.01$ at time-averaged over $t=5000-10000\,t_g$, $10000-15000\,t_g$, $15000-20000\,t_g$, and $25000-30000\,t_g$, respectively. The figure clearly demonstrates that with time, the matter in the high-density thin disc spreads to the off-equatorial plane, and the disc thickness increases. It also shows that, the flow forms a mini-torus-like structure during its evolutionary phase close to the black hole. 

\begin{figure}
    \centering
    \includegraphics[scale=0.4]{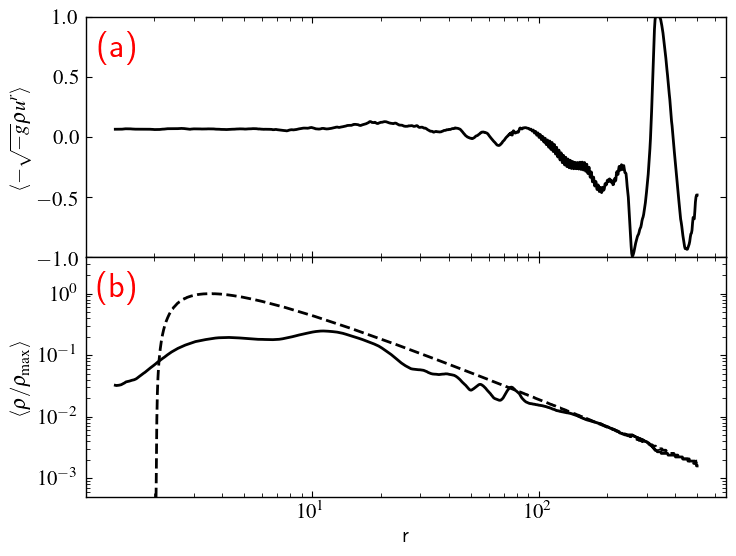}
    \caption{(a) Radial plot of vertically integrated time-averaged ($t=25000-30000\,t_g$) accretion flux $\langle -\sqrt{-g}\rho u^r \rangle$. (b) Radial plot of vertically integrated time-averaged ($t=25000-30000\,t_g$) density profile ($\langle \rho/
    \rho_{\rm max} \rangle$, solid line) and initial vertically integrated density profile (dashed line) for the reference model. }
    \label{fig:inflow-equ}
\end{figure}

Before studying more detail about the flow properties, we would like to check the inflow equilibrium of the accretion flow. To study it, in Fig.~\ref{fig:inflow-equ}, we plot vertically integrated time-averaged (a) accretion flux $\langle -\sqrt{-g}\rho u^r \rangle$ and (b) density profile ($\langle \rho/\rho_{\rm max}\rangle$, solid line) for the reference model. In panel (b), we also plot the vertically integrated initial density profile with the dashed line. We take an average within simulation time $t=25000-30000\,t_g$. In the panel Fig.~\ref{fig:inflow-equ}a, we observe that the vertically integrated time-averaged accretion flux is roughly flat within radius $r\le 100\,r_g$. It suggests that the flow reaches an inflow equilibrium state at least up to radius $r\le 100\,r_g$ \citep[see,][]{Narayan-etal2012}. Further, in the panel Fig.~\ref{fig:inflow-equ}b, we see that by the time $t=25000-30000\,t_g$, the vertically integrated density decreases significantly from the initial value. This indicates the presence of strong disc winds in the accretion flow. We will discuss more detail in section 6. However, at higher radii, the density profile looks quite similar to the initial profile. Note that, after radius $r>150\,r_g$, the density profiles look exactly the same as the initial density, suggesting a not properly evolved accretion disc in the outer part. Accordingly, in the later part of the paper, we can safely use simulation results within $t=25000-30000\,t_g$ to study physics within radius $r\lesssim100r_g$.

\begin{figure}
    \centering
    \includegraphics[scale=0.35]{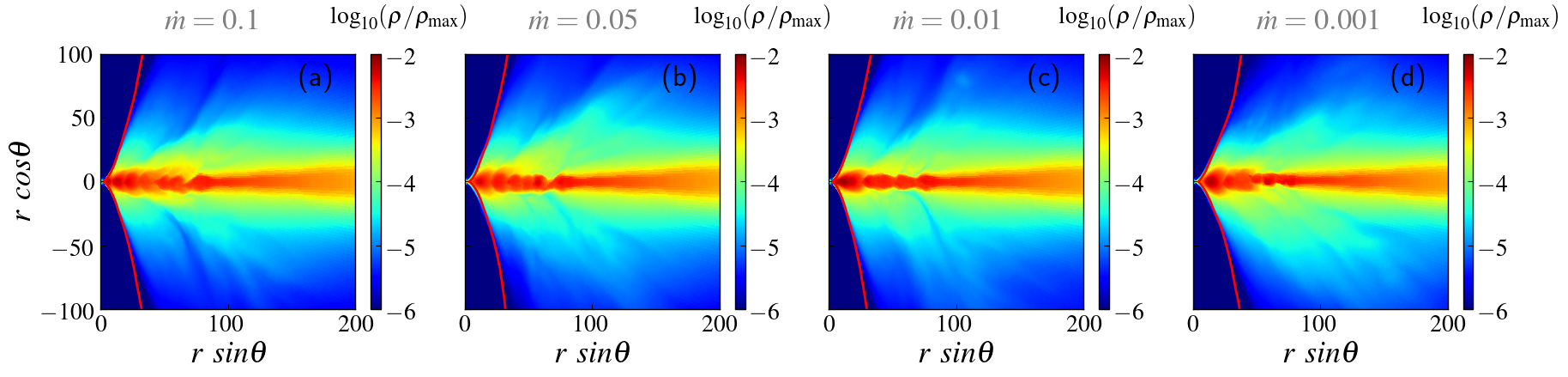}
    \includegraphics[scale=0.35]{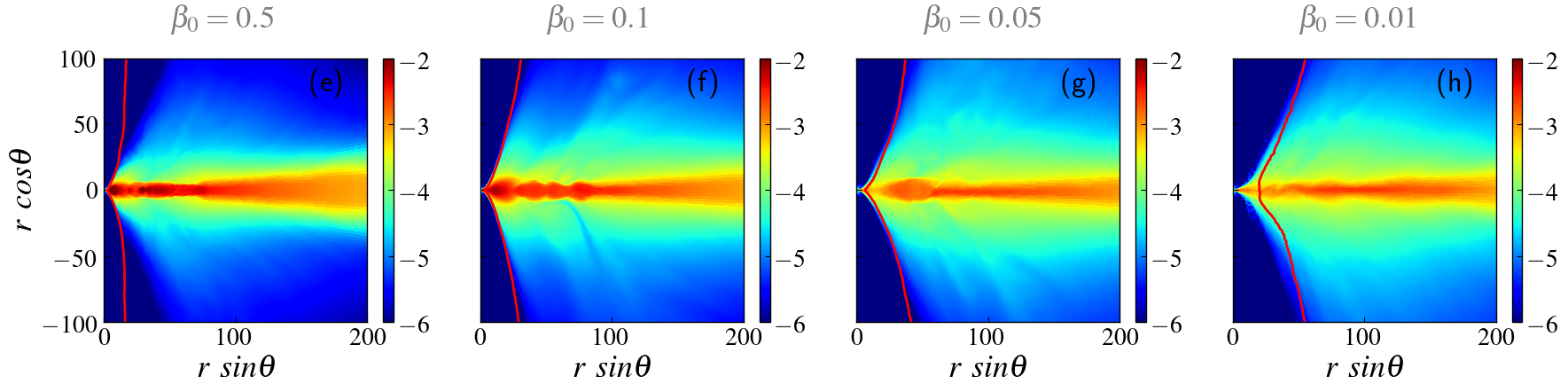}
    \caption{Distribution of time-averaged ($t=25000-30000\,t_g$, for panel (h), $t=10000-15000\,t_g$) logarithmic normalised density 
    $(\log_{10}(\rho/\rho_{\rm max}))$ on the poloidal plane for simulation models with different accretion rates 
    (top) and for different magnetic field strengths through $\beta_0$ (bottom). The solid red line corresponds to $\sigma=1$ contour. 
    }
    \label{fig:density}
\end{figure}

The structure of the disc may also depend on the other flow parameters. Here, in this section, we study the disc structure with input accretion rate and magnetic field strength (input plasma-$\beta$ parameter, i.e., $\beta_0$). In Fig.~\ref{fig:density}, we show the time-averaged logarithmic normalised density distribution of $(\log_{10}(\rho/\rho_{\rm max}))$ for the cases with different input accretion rates $(\dot{m})$ and different input plasma-$\beta$ parameters $\beta_0$. The time averaging is computed over a simulation time $t=25000-30000\,t_g$ for all the models except for model \texttt{G}, which is performed within $t=10000-15000\,t_g$, and the values of $\dot{m}$ and $\beta_0$ are marked on the top of the panels. The red solid contours represent the boundary of $\sigma=b^2/\rho=1$. For the upper panels, we consider $\beta_0=0.1$, and for the lower panels, we consider $\dot{m}=0.01$. In the upper panels, we study the role of accretion rate on the disc structure (Fig.~\ref{fig:density}a-d). 

With the increase in the input accretion rate, the radiative cooling efficiency increases, which leads to a cooler thin disc. As a result, we observe a thinner high-density region for higher accretion rate cases (Fig.~\ref{fig:density}a-d). In the lower panels of Fig.~\ref{fig:density}, we study the role of the magnetic field strength on the disc structure. With the increase in magnetic field strength (Fig.~\ref{fig:density}e-h), the matter in the thin disc reduces drastically due to the strong disc winds (which will be discussed in upcoming sections). By comparing panels of Fig.~\ref{fig:density}e and f, we see that due to the higher magnetic pressure in the disc, the disc thickness increases (see section 5.3 (Fig.~\ref{fig:mag-energy}) for comparison of magnetic field strengths in different models). Moreover, in panels Fig.~\ref{fig:density}g and h, the density close to the equatorial plane is reduced by orders of magnitude.

\begin{figure}
    \centering
    \includegraphics[scale=0.4]{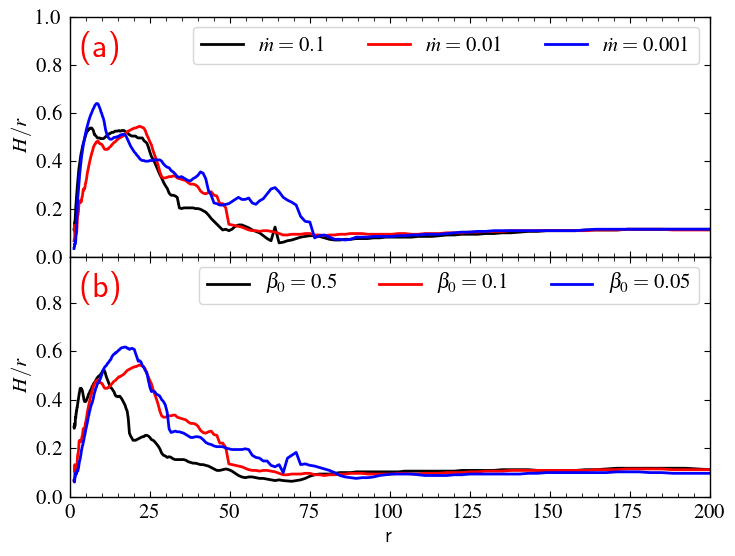}
    \caption{Plot of time-averaged ($t=25000-30000\,t_g$) aspect ratio $(H/r)$ for different models: (a) varying accretion rates and (b) varying magnetic field strengths.}
    \label{fig:height}
\end{figure}

To make a more quantitative comparison between the disc structure, in Fig. \ref{fig:height}, we plot the time-averaged ($t=25000-30000\,t_g$) aspect ratio of the disc for different simulation models: (a) varying accretion rates and (b) varying magnetic field strengths. Here, we calculate the disc height $(H)$ by tracking the location of the height at a given radius where the density drops $\exp(-2)$ of the maximum value at that radius. The figure suggests that close to the black hole, $r<50r_g$, the aspect ratio is significantly greater than $0.1$. This indicates that the evolving disc close to the black hole is thick rather than thin due to the excess of magnetic pressure. But far from the black hole $(r>75r_g)$, the value of the aspect ratio is less than $H/r\sim0.1$, suggesting a thin disc in this region. Note that this value is almost double the initial value of the aspect ratio at radius $(r>75r_g)$, i.e., $H/r|_0\sim0.05$. The dependency of the accretion rate on the aspect ratio is not very significant due to the strong magnetic pressure close to the black hole. However, we observed a slightly lower value of the aspect ratio for the higher values of the accretion rates. On the other hand, for a lower value of the magnetic field strength (higher $\beta_0$), the aspect ratio decreases significantly (see the black solid line in Fig.~\ref{fig:height}b). The panel Fig.~\ref{fig:height}b also shows that with the increase of magnetic field strength (lowering $\beta_0$), the aspect ratio increases.

In summary, we found that accretion rate and magnetic field strength play a vital role in the evolution of the thin-disc structure. In the following sections, we investigate their effects in great detail and their applications in understanding the physics around BH-XRBs. 

\section{Properties of temperatures}
\begin{figure}
    \centering
    \includegraphics[scale=0.35]{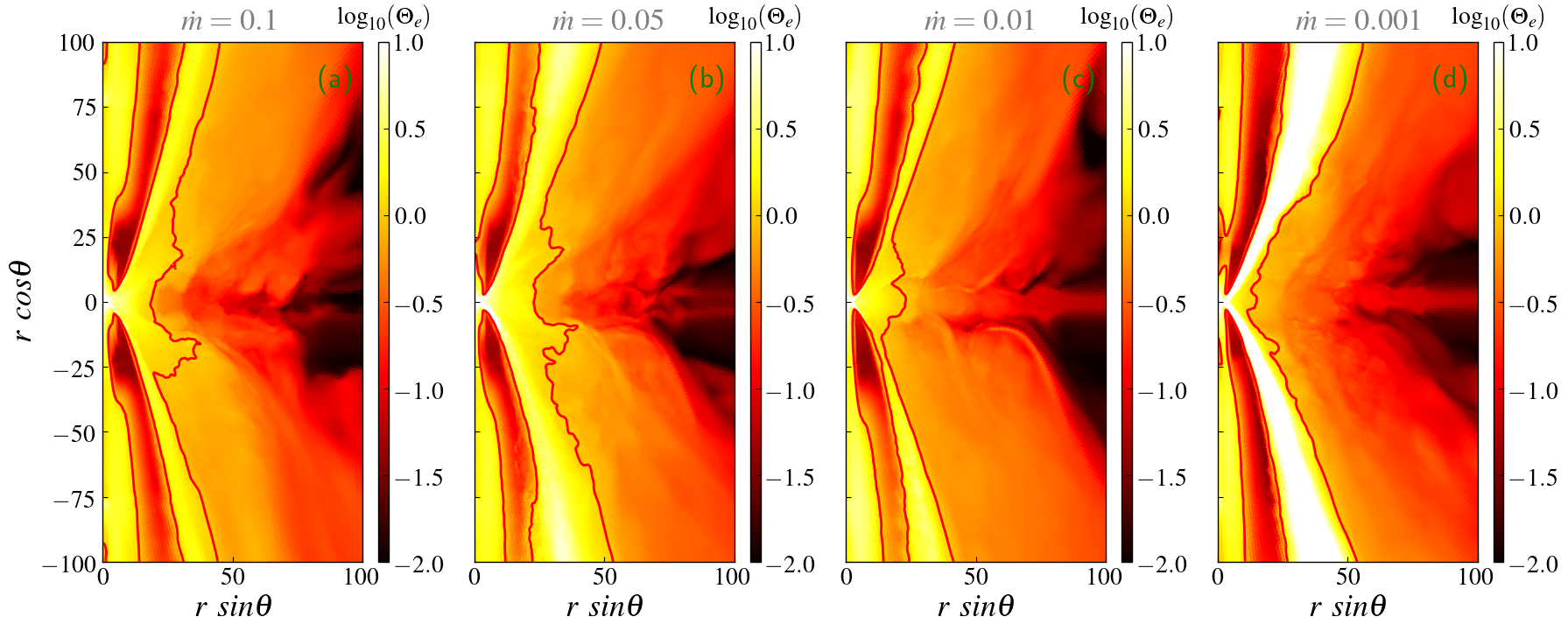}
    \includegraphics[scale=0.35]{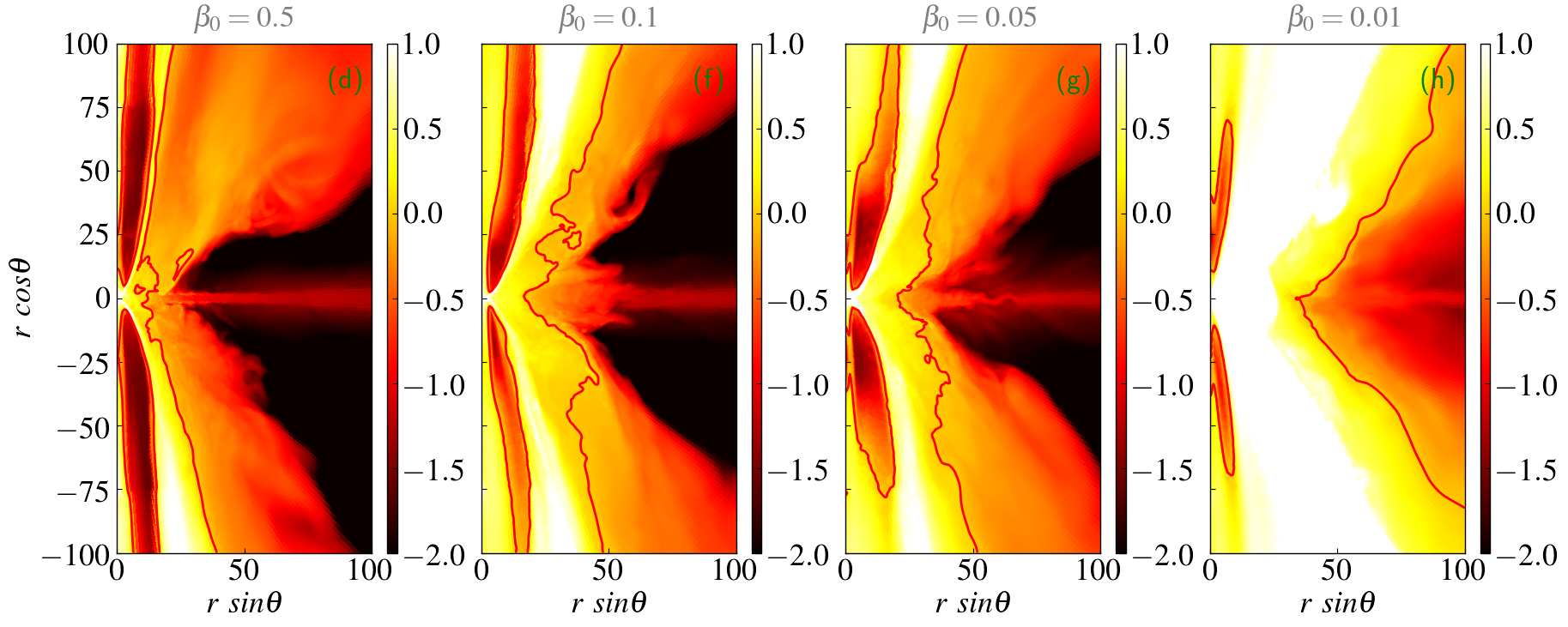}
    \caption{Same as Fig. \ref{fig:density}, except showing the dimensionless electron temperature 
    $\log_{10}(\Theta_e)$. The solid red line corresponds to $\Theta_e=1$ contour.}
    \label{fig:temp_e}
\end{figure}

The radiative properties of an accretion disc have a direct correlation with the temperature of the fluid, particularly the electron temperature. In this section, we study the properties of the electron and ion temperatures of the accretion flow. To do that, we plot logarithmic dimensionless time-averaged (over $t=25000-30000\,t_g$, for model \texttt{G}, $t=10000-15000\,t_g$) electron temperature $\Theta_e=k_{\rm B}T_e/m_e c^2$ in Fig.~ \ref{fig:temp_e} follows the same fashion as Fig.~ \ref{fig:density}. The solid red line in the figure depicts the boundary for $\Theta_e=1$. In all the panels, we see that electrons are hotter in the jet-sheath/corona region due to the strong sub-grid turbulent heating in these regions ($\Theta_e\gtrsim10$, see the bright white-yellow region in Fig.~\ref{fig:temp_e}). In other regions of the simulation domain, the electron temperature is always smaller ($\Theta_e\lesssim1$). With the increase in the input accretion rate $(\dot{m})$, the overall density of the flow increases, which results in higher cooling efficiency for all the radiative processes. Accordingly, electrons are cooler for higher accretion rate models (see upper panels of Fig.~\ref{fig:temp_e}). For models with a lower accretion rate ($\dot{m}=0.01, 0.001$), the electrons close to the equatorial plane but far from the black hole are comparatively hotter as compared to the models with a higher input accretion rate $(\dot{m}=0.1, 0.05)$ (compare panels Fig.~\ref{fig:temp_e}(a,b) and Fig.~\ref{fig:temp_e}(c,d)). For example, in the case with $\dot{m}=0.01$, electrons at the equatorial plane have a temperature of $\Theta_e\sim0.3$. However, for the case with $\dot{m}=1.0$, electrons at the equatorial plane have a temperature of the order of $\Theta_e\sim0.03$, which is around ten times smaller. Such a low electron temperature around the equatorial plane essentially suggests the importance of radiative cooling due to the optically-thick black body component in these ranges of accretion rates ($\dot{m}\gtrsim0.5$). 

Although, with the increase in magnetic field strength, the emission due to synchrotron radiation also increases. However, in Fig.~\ref{fig:temp_e}, it is shown that electrons become hotter for models with a lower initial $\beta_0$ (stronger magnetic field). This essentially suggests that the turbulent heating present in the flow is much more efficient than that of radiative cooling. Consequently, we see that the area under the $\Theta_e=1$ contour increases with the increasing magnetic field strength (lowering $\beta_0$). For higher initial $\beta_0$ cases, the electrons away from the equatorial plane surrounding the corona region are cooled ($\Theta_e\lesssim0.01$, see the dark black region of the panels Fig.~\ref{fig:temp_e}e, \ref{fig:temp_e}f, and \ref{fig:temp_e}g). With the increase in magnetic field strength $(\beta_0=0.01)$, the electron temperature in this region increases by orders of magnitude ($\Theta_e\sim0.3$, see the reddish region of the panel in Fig.~\ref{fig:temp_e}h).
\begin{figure}
    \centering
    \includegraphics[scale=0.6]{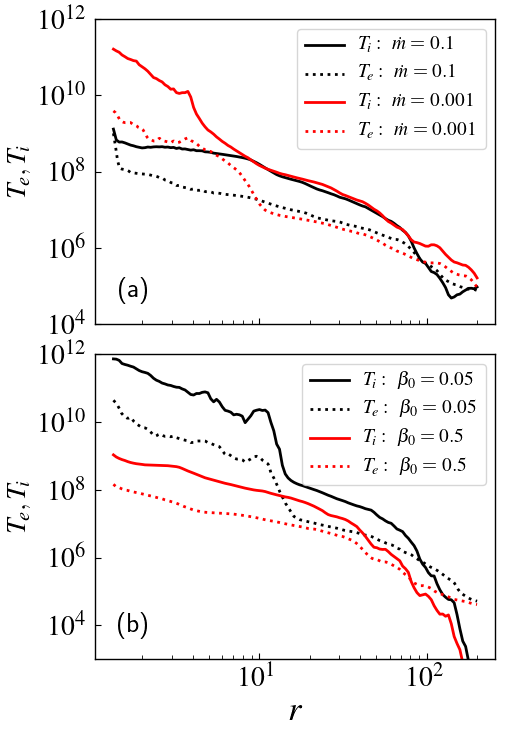}
    \caption{The comparison of dimensional electrons and ions temperatures ($T_i, T_e$, in Kelvin) for 
    different (a) input accretion rates (upper panel) and (b) input plasma-$\beta$ parameters ($\beta_0$, lower panel) 
    is shown on the equatorial plan. The solid lines represent ion temperature, whereas the dotted lines represent 
    electrons temperature.}
    \label{fig:temp-1D}
\end{figure}

To study the temperature profiles in the thin disc, we compare the time-averaged temperatures of electrons and ions ($T_e, T_i$, in Kelvin) for various accretion rates (upper panel) and initial plasma-$\beta$ (lower panel) parameters on the equatorial plane in Fig.~\ref{fig:temp-1D}. The time averaging for the temperatures is taken at simulation times between $t=25000-30000\,t_g$. In the upper panel, we fixed the initial plasma-$\beta$ at $\beta_0=0.1$, while in the lower panel, we fixed the input accretion rate at $\dot{m}=0.01$. In the figure, the solid and dotted lines represent ion and electron temperatures, respectively. In Fig.~\ref{fig:temp-1D}a, for the lower accretion rate (red lines), far from the black hole ($r>30\,r_g$) electrons are hotter than ions, which is due to inefficient electron cooling mechanisms. With the increase in accretion rate, the electrons cool efficiently due to the optically thick component of the radiation, and as a result, the electrons become cooler than the ions (see the black lines). However, in the intermediate radii ($2\,r_g\lesssim r \lesssim 30\,r_g$) the temperatures of the electrons and ions do not significantly depend on the accretion rates. On the other hand, in the region near the black hole $(r\lesssim 2\,r_g)$, the temperatures of electrons and ions are lower for higher accretion cases due to higher Bremsstrahlung and Coulomb heating processes. It is interesting to note that the ion and electron temperatures close to the black hole reach as high as $T_i\sim10^{11}$K and $T_e\sim10^{9}$K.

Figure~\ref{fig:temp-1D}b shows that for the weak magnetic field case ($\beta_0=0.5$), the temperatures of electrons and ions are much smaller close to the black hole as compared with the high magnetic field case ($\beta_0=0.05$). The maximum ion temperature is of the order of $T_i\sim10^{12}$K and $T_i\sim 10^{9}$K for $\beta_0=0.05$ and $\beta_0=0.5$, respectively. Similarly, the maximum electron temperature is of the order of $T_e\sim10^{11}$ and $T_e\sim 10^{8}$K for $\beta_0=0.05$ and $\beta_0=0.5$, respectively. These facts essentially suggest the importance of the turbulent heating process in determining the temperature of both the electrons and ions. The impacts of the initial plasma-$\beta$ parameters can be seen throughout the length scale of the accretion disc. Even far from the black hole ($r>20\,r_g$ for $\beta_0=0.5$ and $r> 30\,r_g$ for $\beta_0=0.05$), electrons remain hotter than ions for both the magnetic field limits. 

\begin{figure}
    \centering
    \includegraphics[scale=0.35]{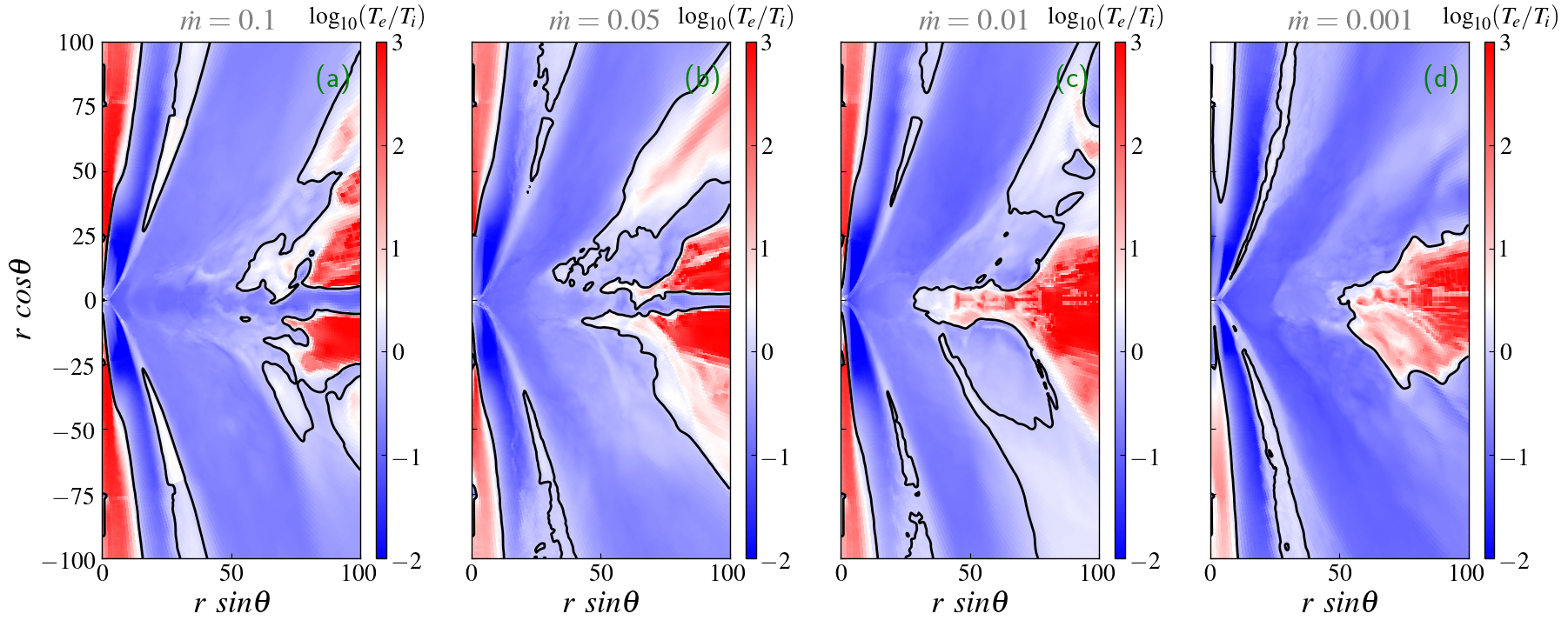}
    \includegraphics[scale=0.35]{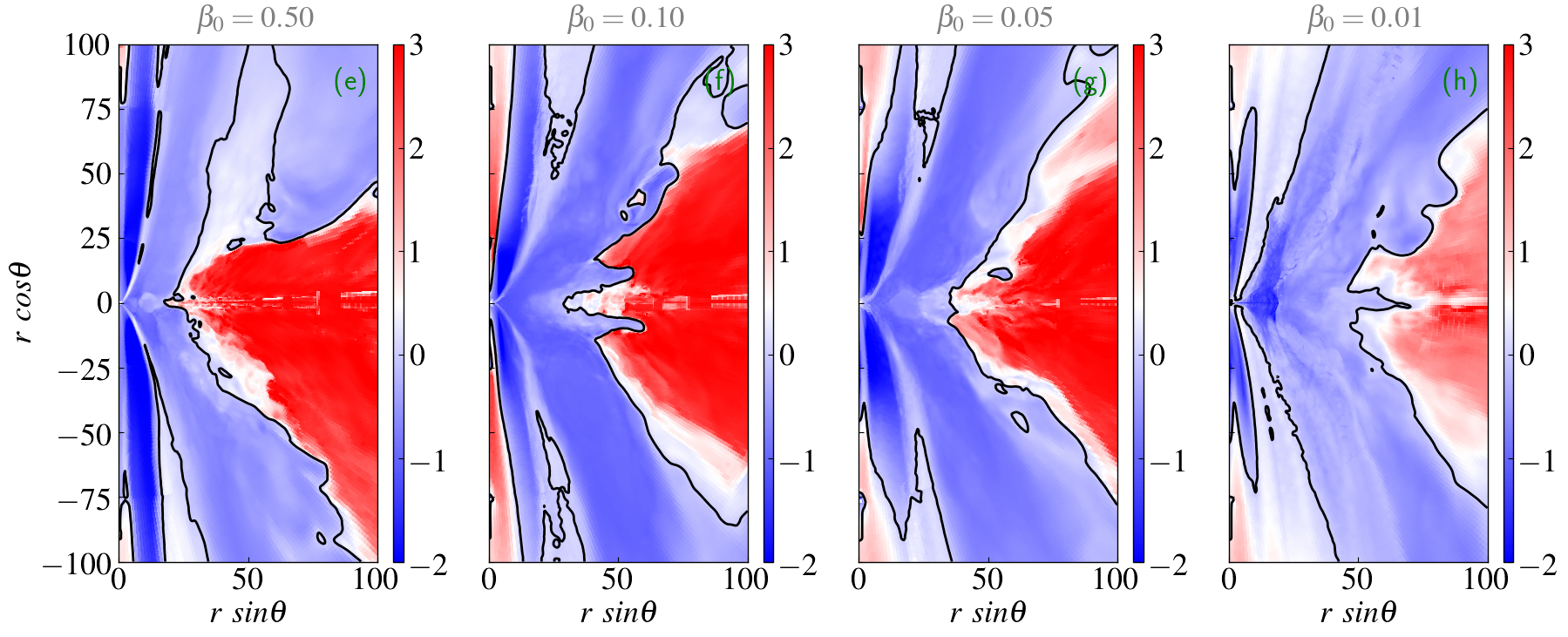}
    \caption{Same as Fig. \ref{fig:density}, except showing the ratio between dimension-full electron and 
    proton temperatures in $\log_{10}(T_e/T_i)$. A solid black line corresponds to $T_e/T_i=1$ contour.}
    \label{fig:ratio}
\end{figure}

In a similar spirit as in Fig.~\ref{fig:density}, Fig.~ \ref{fig:ratio} shows the distribution of the time-averaged (over $t=25000-30000\,t_g$, for model \texttt{G}, $t=10000-15000\,t_g$) ratio between dimensional electron and ion temperatures as $\log_{10}(T_e/T_i)$, where the boundary of $T_e/T_i=1$ is represented as a solid black curve. The ratio $T_e/T_i$ essentially depicts the distribution of thermal energy in the fluid over electrons and ions. In the corona region, the thermal energy of the electron is always lower than that of the ion thermal energy ($T_e/T_i<1$). The effects of the increasing radiative cooling with the accretion rate can be seen in the upper panels of figure~\ref{fig:ratio}. For a higher accretion rate, the thermal energy of electrons is lower than that of ions in the thin-disc region (see the blue region in Fig.~\ref{fig:ratio}a and \ref{fig:ratio}b). However, for a lower accretion rate, the thermal energy of the electron remains higher than that of the ions in the thin-disc region. Away from the corona region, the thermal energy of ions is always higher than that of electrons (see the red region in Fig.~\ref{fig:ratio}a-\ref{fig:ratio}h).

The ratio $T_e/T_i$ shows a non-monotonic behaviour with the increase in magnetic field strength (lowering $\beta_0$, Fig.~\ref{fig:ratio}e-\ref{fig:ratio}h). Initially, with the increase of the magnetic field strength, the emission due to the Synchrotron radiation increases. The electron loses its thermal energy. As a result, the area with $T_e/T_i<1$ in the corona region increases, and also the ratio $T_e/T_i$ decreases (see the darker blue colour in panel Fig.~\ref{fig:ratio}g as compared to panel \ref{fig:ratio}f). However, with a further increase in the magnetic field strength, the turbulent heating efficiency also increases, leading to an increase in the ratio of $T_e/T_i$ in the corona. With this, the ratio of temperatures becomes greater than unity ($T_e/T_i>1$) in some parts of the corona region (see Fig.~\ref{fig:ratio}h). 

\subsection{Temperatures with heating prescriptions}
\begin{figure}
    \centering
    \includegraphics[scale=0.38]{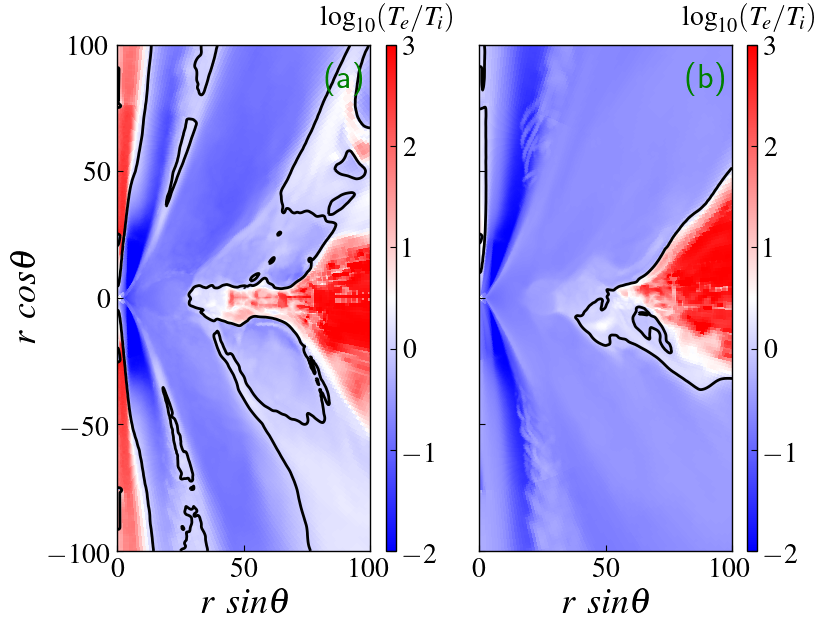}
    \caption{Distribution of time-averaged temperatures ratio ($\log_{10}(T_e/T_i$)) for (a) turbulent heating (model \texttt{C}) 
    and (b) reconnection heating (model C-R) for electrons. The black solid line represents the boundary of 
    $T_e/T_i=1$.}
    \label{fig:comp-heating}
\end{figure}

In this section, we study the impact on the ratio of temperatures ($T_e/T_i$) with different heating prescriptions for electrons. Accordingly, in Fig. \ref{fig:comp-heating}, we show the time-averaged (over $t=25000-30000\,t_g$) distribution of $T_e/T_i$ for two heating prescriptions (a) turbulent heating \citep{Howes2010,Howes2011} and (b) reconnection heating \citep{Rowan-etal2017}, i.e., for model \texttt{C} and model \texttt{C-R}, respectively. For both models, the input accretion rate and the initial magnetic flux strength are fixed to $\dot{m}=0.01$ and $\beta_0=0.1$, respectively. The figure essentially suggests that the electrons in the funnel region are mostly cooler than the ions for the reconnection heating model, implying the inefficiency of reconnection heating in the highly magnetised regions. Whereas, for the turbulent heating model, we observe a significant area with $T_e/T_i>1$ around the funnel region. Also, the area with $T_e/T_i<1$ is significantly bigger for the reconnection heating model as compared to the turbulent heating model. However, the region with $\beta\ll1$ above the thin remains $T_e/T_i>1$ for both heating prescriptions.

\subsection{Correlation of \texorpdfstring{$T_{e}/T_{i}$}{} with plasma-\texorpdfstring{$\beta$}{}}
In this section, we focus on the correlation between the ratio of temperatures ($T_e/T_i$) and the flow plasma-$\beta$ profile. In previous study, \cite{Dihingia-etal2023} proposed a $R-\beta$ ansatz that provides the trend of $T_e/T_i$ with flow plasma-$\beta$ in the presence of radiative cooling and heating, which is given by,
\begin{align}
    \frac{T_i}{T_e}= \frac{1}{1+\beta^2}R_{1} +
    \frac{\beta^2}{1+\beta^2} R_{2} + \frac{\beta}{\beta+\beta_{{\rm
          br}}} R_{3} \,,
    \label{eq:rbeta}
\end{align}
where $R_{1-3}$ are constant determining the nature of $T_i/T_e$ in different ranges of plasma-$\beta$. For, $\beta\ll1$ the ratio becomes $T_i/T_e=R_1$, whereas for $\beta\gg1$ the ratio becomes $T_i/T_e=R_2+R_3$. Note that this ansatz recovers the form of the $R-\beta$ relation provided by \cite{Moscibrodzka-etal2016} when $R_3=0$. In that case, $R_1=R_{\rm low}$ and $R_2=R_{\rm high}$, i.e., 
\begin{align}
    \frac{T_i}{T_e}= \frac{R_{\rm low}}{1+\beta^2} + \frac{R_{\rm high}
      \beta^2}{1+\beta^2}\,.
    \label{eq:rbetaorg}
\end{align}
Recently, \cite{Meringolo-etal2023} also proposed such a relation between the temperature ratio, 
the plasma-$\beta$, and magnetisation $(\sigma)$ with kinetic Particle-In-Cell (PIC) simulations, which is given by,
\begin{align}
    \frac{T_i}{T_e}=\left[ t_0 + t_1  \sigma^{\tau_1} \tanh\left[t_2 \beta \sigma^{\tau_2}\right] +  
    t_2  \sigma^{\tau_3} \tanh\left[t_3 \beta^{\tau_4} \sigma\right]\right]^{-1}\,.
    \label{eq:Meringolo-etal2022}
\end{align}
where $t_0 = 0.4$, $t_1 = 0.25$, $t_2 = 5.75$, $t_3 = 0.037$, $\tau_1 = -0.5$, $\tau_2 = 0.95$, 
$\tau_3 = -0.3$, and $\tau_4 = -0.05$. In \cite{Dihingia-etal2023}, we consider the accretion flow to be geometrically thick (torus), and such 
accretion flows are applicable around low-luminous AGNs. However, here we study a geometrically thin disc, 
and therefore, we want to study the possibility of a similar correlation between the ratio of temperatures and 
the plasma-$\beta$ profile.

\begin{figure}
    \centering
    \includegraphics[scale=0.3]{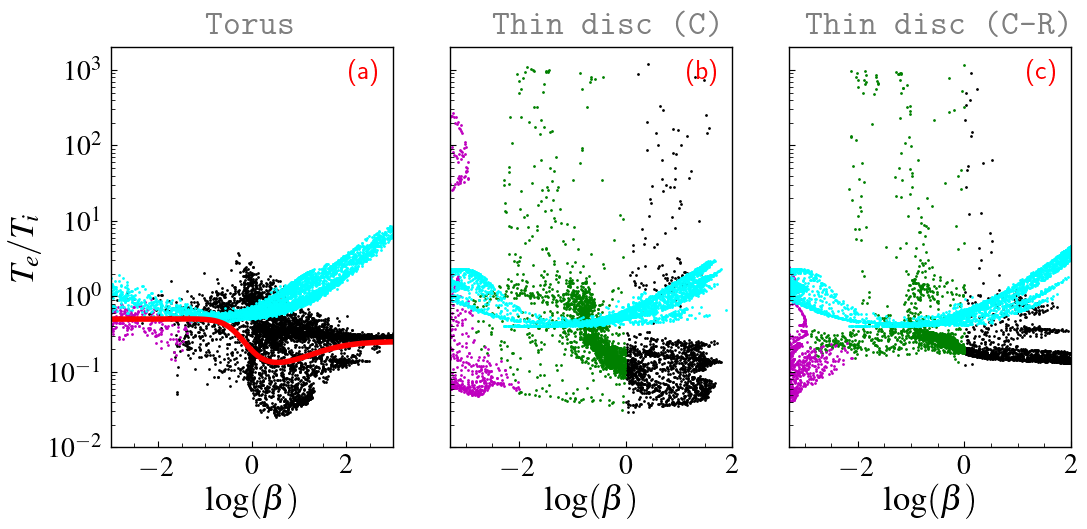}
    \caption{Distributions of the time-averaged temperatures ratio $T_e/T_i$ as a function of 
    plasma-$\beta$ for geometrically (a) thick, (b) thin accretion flow (model \texttt{C}), and (c) thin accretion flow 
    (model \texttt{C-R}). Each dot refers to a 
    cell in the computational domain, with the black dots referring to the region with $\sigma<1$ 
    and the magenta dots to the region with $\sigma>1$ (roughly representative of the jet). The green dots in panels (b) and (c) correspond to the region $\beta<1$ and $\sigma<1$. Finally, cyan dots represent temperature ratios using Eq.~\ref{eq:Meringolo-etal2022}.
    The solid red curves correspond to $R-\beta$ relations.}
    \label{fig:correlation}
\end{figure}

In order to fully understand the differences in the correlation in these two scenarios (torus and thin disc), in Fig.~\ref{fig:correlation}, we collect the time-averaged ratio of the temperatures $T_e/T_i$ from all the grid cells in the computational domain with $r<100\,M$. To avoid inconsistencies due to the SMR grids, we converted them to uniform grids and performed the calculations. In the plot, we reduced the plotting number of points to make the plot presentable, considering more points do not change the morphology of the plots. The distribution for torus is shown in the panel Fig.~\ref{fig:correlation}a and the distribution for thin-disc is shown in Fig.~\ref{fig:correlation}b and \ref{fig:correlation}c. Panel~\ref{fig:correlation}b shows the correlations of temperatures for model \texttt{C} and in panel \ref{fig:correlation}b shows the correlations of temperatures for model \texttt{C-R}. The black and magenta dots refer to the ratio collected from the region with $\sigma<1$ and $\sigma>1$ (roughly representative of the jet) region, respectively. The green dots in panels (b) and (c) correspond to values with $\beta<1$ and $\sigma<1$, roughly representing the disc-wind region. For the torus, we use the model \texttt{TRC-A} of the \cite{Dihingia-etal2023} and for the thin disc, we consider our reference model (model \texttt{C}). The panels essentially suggest that the ratio $T_e/T_i$ behaves in quite a different manner. For the torus case, it is possible to fit all the data roughly with Eq.~\ref{eq:rbeta}, which is shown by the red solid line. However, for the thin disc case, some data points have a much larger temperature ratio, $T_e/T_i\gg1$ in the range between $10^{-3}\leq \beta\leq 10^{-1}$. Due to the presence of these data points, it is impossible to fit the data using Eq.~\eqref{eq:rbeta}. Consequently, we avoid doing it. The cyan dots represent the temperature ratios calculated using Eq.~\eqref{eq:Meringolo-etal2022}. We observe that the cyan dots overlap with the simulated temperature ratios in the simulation domain for the highly magnetised region $(\beta\ll1)$. However, in the low-magnetised region, the relation provided by \cite{Meringolo-etal2023} does not provide good agreement. 

\begin{figure}
    \centering
    \includegraphics[scale=0.38]{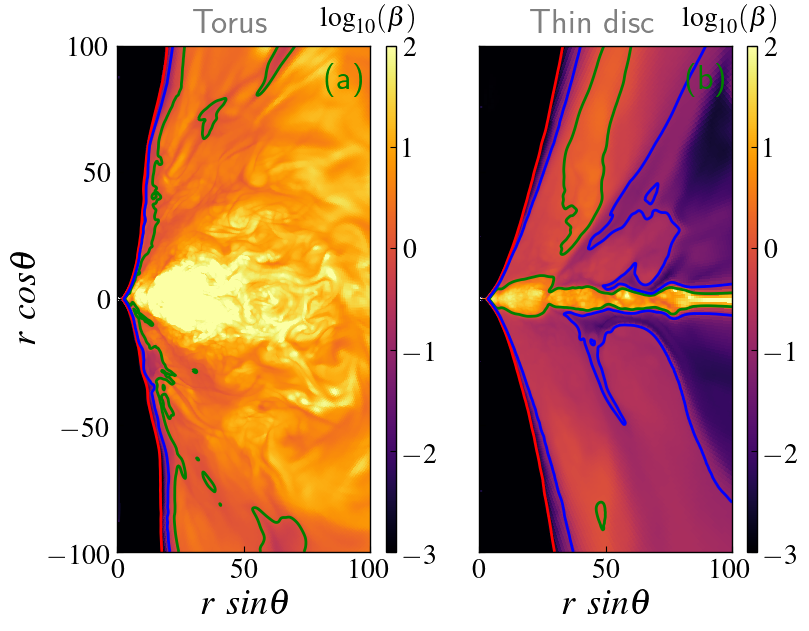}
    \caption{Distribution of flow plasma-$\beta$ profile for torus (a) and thin-disc (b) simulation. 
    The red, green, and blue lines represent the boundary of $\sigma=1$, $\beta=1$, and $\beta=0.1$, 
    respectively.}
    \label{fig:comp-beta}
\end{figure}

Figure~\ref{fig:comp-beta} shows the plasma-$\beta$ distribution for both torus and thin-disc models, where the red, green, and blue lines represent the boundary of $\sigma=1$, $\beta=1$, and $\beta=0.1$, respectively. For the torus model (Fig.~\ref{fig:comp-beta}a), the plasma-$\beta$ is mostly greater than that of unity except in the relativistic jets ($\sigma>1$ region), i.e., the flow is mostly gas pressure dominated. For the thin-disc model (Fig.~\ref{fig:comp-beta}b), we see that around the equatorial plane, the flow is gas pressure dominated $\beta\gg1$, as the flow moves away from the equatorial plane, it becomes magnetic pressure dominated ($\beta<1$, see the dark region of the Fig.~\ref{fig:comp-beta}b). However, around the rotation axis (jet and corona) of the black hole, the distribution of plasma-$\beta$ is quite similar to the torus setup. Thus, the scenario of plasma-$\beta$ distribution is completely different for the thin-disc due to the presence of the low-$\beta$ region between the thin-disc and the corona. This low-$\beta$ region essentially represents the \cite{Blandford-Payne1982} dominated disc-wind (see next section for details). In summary, in the case of a magnetised thin-disc, the simplified $R-\beta$ prescription cannot reproduce the trend of the electron-to-ion temperature ratio ($T_i/T_e$) in terms of flow plasma-$\beta$. Accordingly, in such a scenario, a two-temperature framework is unavoidable for a proper description of electron temperature.

\subsection{Components of the outflow}
\begin{figure}
    \centering
    \includegraphics[scale=0.40]{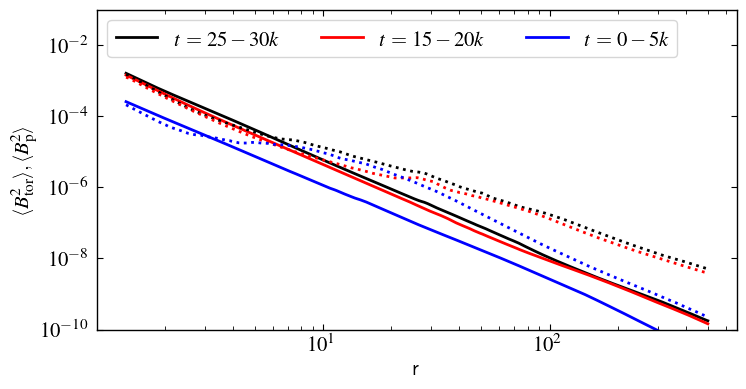}
    \caption{Plot for the time evolution of time-averaged, vertically integrated strength of the different magnetic field components ($\langle B^2_{\rm p}\rangle$ (solid) and $\langle B^2_{\rm tor}\rangle$ (dotted)). Black, red, and blue correspond to time-averaged between $t=25000-30000\,t_g$, $15000-20000\,t_g$, and $0-5000\,t_g$, respectively.}
    \label{fig:mag-energy-time}
\end{figure}
In this section, we study the components of outflow. The outflow (jet/winds) depends on the strengths of the magnetic fields. Therefore, before going into further detail, we show the time evolution of the time-averaged, vertically integrated magnetic field strength in terms of $\langle B^2_{\rm p} \rangle$ (solid) and $\langle B^2_{\rm tor} \rangle$ (dotted) in Fig.~\ref{fig:mag-energy-time}. For these calculations, we choose the reference models (model \texttt{C}). The different time ranges are mentioned in the figure. These two components are calculated by following $B^2_{\rm p}=B^r B_r + B^\theta B_\theta$ and $B^2_{\rm tor}=B^\phi B_\phi$. The simulations start with zero toroidal components of the magnetic field. Due to the angular motion of the flow, the toroidal component generates the flow. Accordingly, we initially see a smaller toroidal component (blue dotted line). However, after evolution, the strength increases and reaches a saturated value throughout the radial distances (see black and red dotted curve). Similar to the toroidal components, the poloidal components also increase with time due to the presence of MRI in the flow. Finally, we observe that the strength of the poloidal components saturates at all the radii (see black and red solid lines). Due to the asymmetric nature of the simulation, it is expected that the strength of the magnetic field may decay after a very long simulation \citep{Cowling1933}. However, within the simulation time, we do not observe it.

\begin{figure}
    \centering
    \includegraphics[scale=0.40]{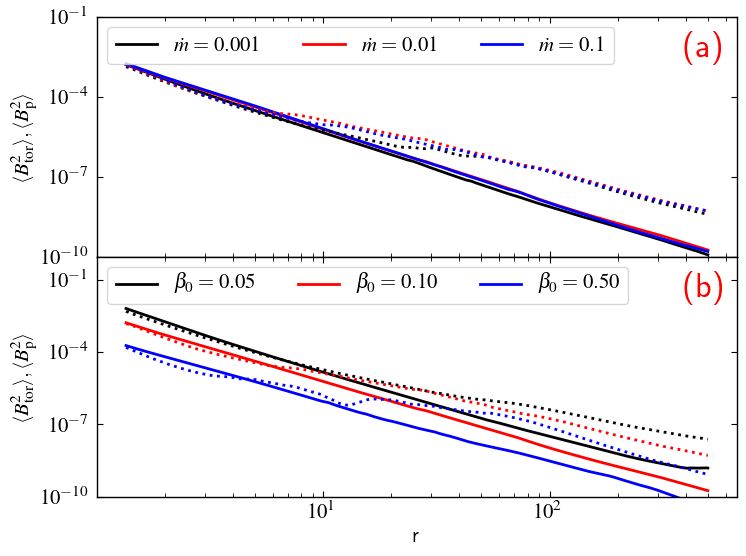}
    \caption{Plot for time-averaged ($t=25000-30000\,t_g$), vertically integrated strength of the different magnetic field components ($\langle B^2_{\rm p}\rangle$ (solid) and $\langle B^2_{\rm tor}\rangle$ (dotted)). The upper and lower panels show the variations for different accretion rates and different initial plasma-$\beta$ parameters.}
    \label{fig:mag-energy}
\end{figure}

Next, we show the time-averaged ($t=25000-30000\,t_g$), vertically integrated magnetic field strength in terms of $\langle B^2_{\rm p} \rangle$ (solid) and $\langle B^2_{\rm tor} \rangle$ (dotted) in Fig.~\ref{fig:mag-energy} for different models. Panels (a) and (b) correspond to variation with accretion rate (with fixed $\beta_0=0.1$) and initial magnetic field strengths ($\beta_0$, with fixed $\dot{m}=0.01$). The panel (a) of Fig.~\ref{fig:mag-energy} suggests that the magnetic field strengths are roughly similar for different initial accretion rates. However, we observe slight differences in the toroidal magnetic field components between $r\sim5-20r_g$. On the contrary, panel (b) of Fig.~\ref{fig:mag-energy} suggests dependency on the initial magnetic field strengths. The strengths of both components increase with the decrease of $\beta_0$. Accordingly, we expect strong dependencies of disc-wind with $\beta_0$, but weak dependencies with the initial accretion rate.

\begin{figure}
    \centering
    \includegraphics[scale=0.42]{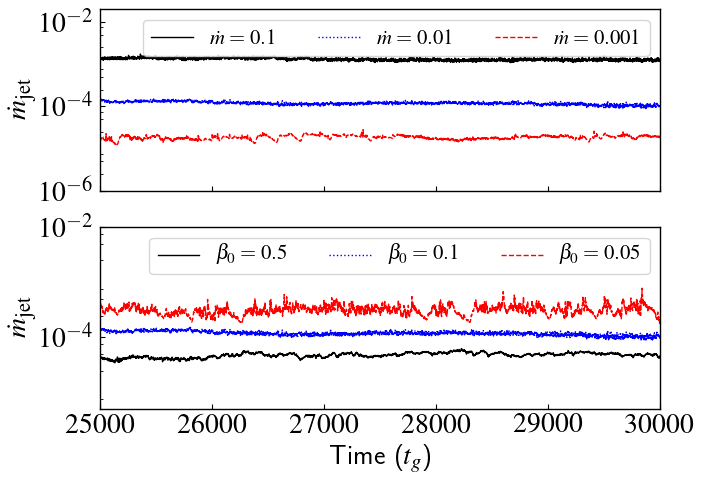}
    \includegraphics[scale=0.42]{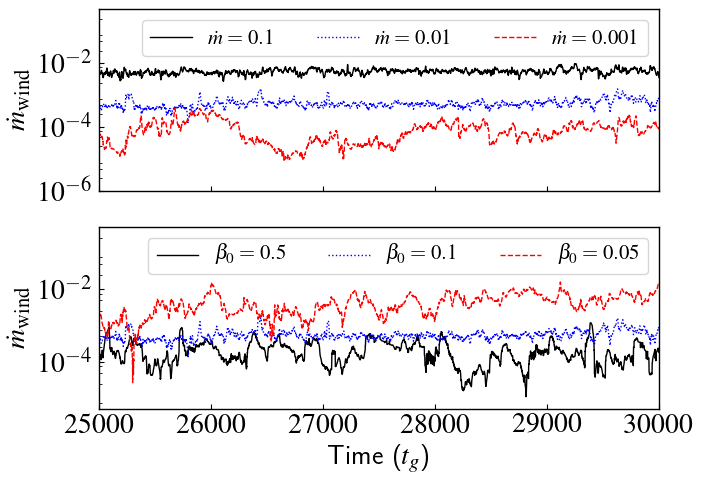}
    \caption{Plots of mass outflow rates through the jet (left) and wind (right) for different simulation models with different values of $\dot{m}$ and $\beta_0$.}
    \label{fig:outflow-rates}
\end{figure}

To understand the properties of outflows, we classify them broadly into two parts: relativistic jets (simply jets) and disc-wind (simply wind). To calculate the mass flux rates through the jet and wind, we follow definitions that are well discussed in many literature \citep[e.g.,][]{Nathanail-etal2020, Dihingia-etal2021, Scepi-etal2023}. We integrate mass flux ($\sqrt{-g}\rho u^r$) over the polar angle at a radius $r=50$ considering all locations with $\sigma>1$ or $\eta>2$, and $-hu_t>1$ to calculate the outflow rate of jets ($\dot{m}_{\rm jet}$). The $\eta$ represents the efficiency factor of the Poynting flux, which is given by $\eta = -(T^r_t - \rho u^r)/(\rho u^r)$. Similarly, we calculate the outflow rate of winds ($\dot{m}_{\rm wind}$) by integrating the same over the polar angle where $\sigma<1$ or $\eta<2$, and $-hu_t>1$. Following these definitions, in Fig.~\ref{fig:outflow-rates}, we plot the mass outflow rates through the jet (left) and wind (right) for different simulation models with different values of $\dot{m}$ and $\beta_0$. The values of $\dot{m}$ and $\beta_0$ are marked on the panels. For the sake of a better presentation, we multiply the variation of the rates with $\dot{m}$ with fractions $10$, $1$, and $0.1$ to mimic the density scaling due to different accretion rates. In the code units, the averaged values are quite similar. However, we observe higher variability amplitudes in both the wind and jet rates for lower accretion rate cases. In the lower panels of Fig.~\ref{fig:outflow-rates}, we see that the mass loss through the jet and wind is correlated with the strength of the magnetic fields. The lower $\beta_0$ (stronger magnetic field) shows larger outflow rates. Although Fig.~\ref{fig:outflow-rates} gives an idea about the mass loss through wind, it cannot determine the mechanism responsible for the wind.

\section{Properties of disc-winds}
\begin{figure}
    \centering
    \includegraphics[scale=0.40]{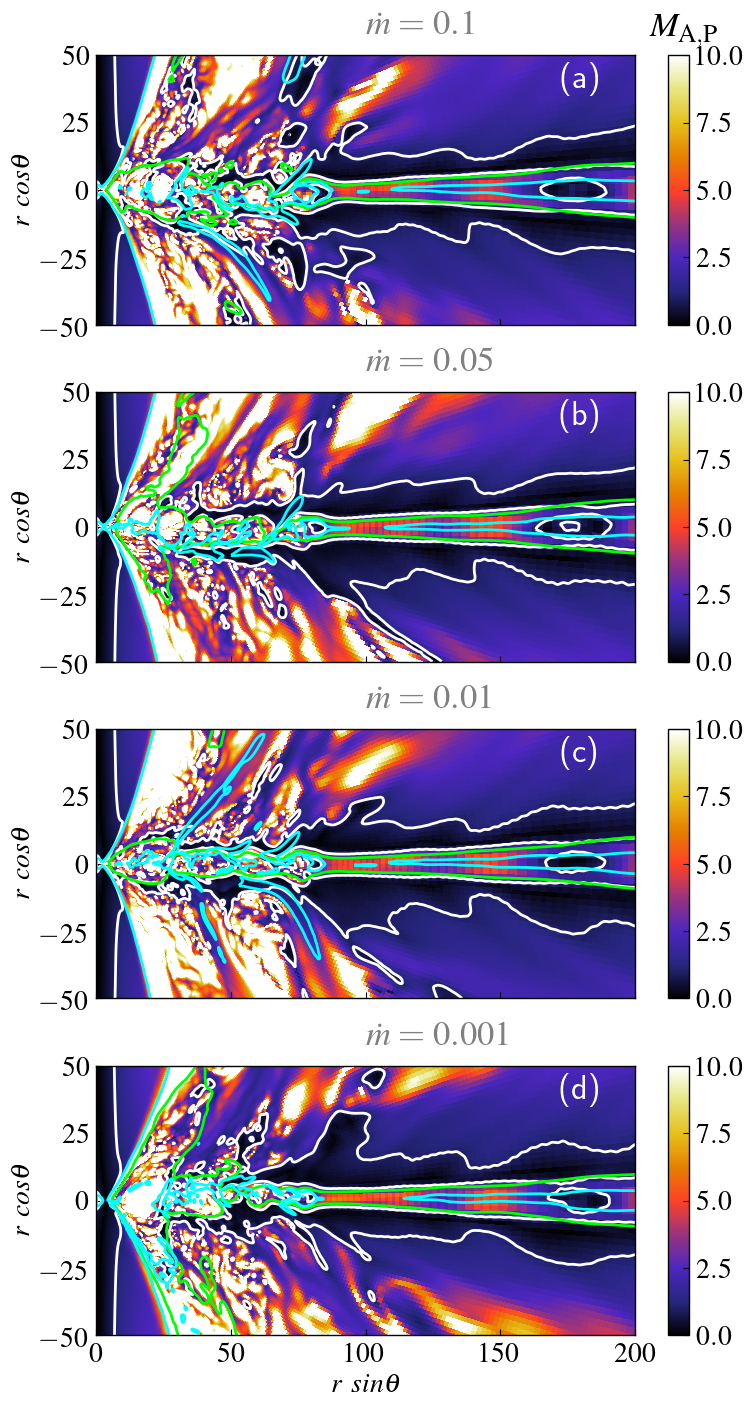}
    \includegraphics[scale=0.40]{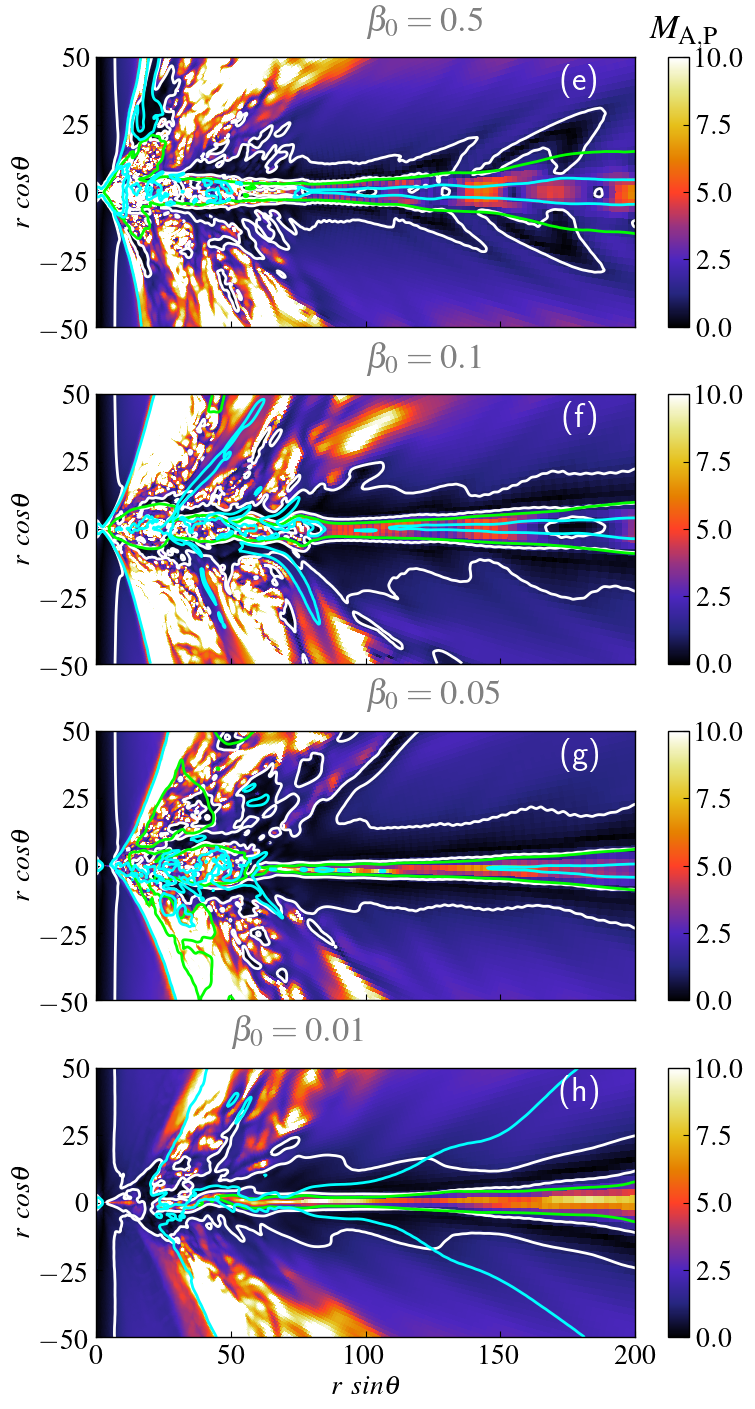}
    \caption{Distribution of poloidal Alfv\'enic Mach number $(M_{\rm A,P})$ for different accretion rates (left panels 
    (e)-(h)) and magnetic field strengths (right panels, (e)-(h)). The white, green, and light blue lines correspond 
    to the boundaries of $M_{\rm A,P}=1$, $\beta=1$, and $B_{\rm tor}/B_p=5$, respectively.}
    \label{fig:disc-wind}
\end{figure}
In this section, we characterize different types of disc winds from the accretion disc and subsequently investigate the role of flow parameters in the launching of disc winds in the presence of radiative cooling and electron heating. To do that, in Fig.~\ref{fig:disc-wind}, we show the distribution of poloidal Alfv\'enic Mach number ($M_{\rm A,P}$) for the different input accretion rates (left panels (e)-(h)) and the magnetic field strengths (right panels, (e)-(h)). The values of input accretion rates $(\dot{m})$ and the initial plasma-$\beta$ parameters ($\beta_0$) are given at the top of each panel. The white, green, and light blue lines correspond to the boundaries of $M_{\rm A,P}=1$, $\beta=1$, and $B_{\rm tor}/B_p=5$, respectively. The poloidal Alfv\'enic Mach number is obtained as $M_{\rm A, p}=u_p/c_a$, where $h=1 + \frac{\Gamma p}{(\Gamma -1)\rho}$, $c^2_{\rm a}=B_p^2/(\rho h + B_p^2)$, $u_p^2 = [u^ru_r + u^\theta u_\theta]_{\rm BL}$, and finally $B_p^2 = [B^rB_r + B^\theta B_\theta]_{\rm BL}$. Following a similar procedure, the toroidal component of the magnetic fields is calculated as $B^2_{\rm tor}=[B^\phi B_\phi]_{\rm BL}$. The subscript BL\footnote{The value of $u_p^2$, $B_p^2$, and $B_{\rm tor}^2$ depends on the coordinates. We calculate them in the physical coordinate system of a black hole, i.e., Boyer-Lindquist coordinates (see \cite{McKinney-Gammie2004,Porth-etal2017}). \label{footnote-01}} indicates that the quantities are calculated at Boyer-Lindquist coordinates. 

In all cases, the region around the rotation axis of the black hole flow is always sub-Alfv\'enic Mach number. In this region, $\beta\ll1$ and $B_{\rm tor}/B_p\ll1$, which indicated the presence of a magnetically driven relativistic jet. Surrounding that, we observe a region with a super-Alfv\'enic Mach number. In this region, $M_{\rm A, p}\gg1$, $\beta\gtrsim1$, and $B_{\rm tor}/B_p\gg1$. It signifies the $B_{\rm tor}$-dominated disc-wind. The launching mechanism for this component of the disc-wind is the gradient of magnetic pressure due to the toroidal component of the magnetic field \citep[e.g.,][]{Vourellis-etal2019, Dihingia-etal2021}. $B_{\rm tor}$ dominated disc-wind is often hot and acts as a hot disc-wind corona around the black hole. Also, plasmoids can be seen in this region \citep{Dihingia-etal2021}.

Far from the black hole, the accretion flow has $M_{\rm A, p}>1$ (super-Alfv\'enic), $\beta\gg1$, and $B_{\rm tor}/B_p\gtrsim1$ around the equatorial plane. However, away from the equatorial plane, flow becomes sub-Alfv\'enic $M_{\rm A, p}<1$, with $\beta\ll1$ and $B_{\rm tor}/B_p>5$. Moreover, flow again becomes super-Alfv\'enic $M_{\rm A, p}>1$ as it moves far from the equatorial plane. This implies that the poloidal magnetic field dominates the region near and above the disc surface, where the wind is centrifugally pushed along the poloidal magnetic field, a characteristic of the \cite{Blandford-Payne1982} disc wind (BP disc-wind) \citep{Qian-etal2018,Vourellis-etal2019,Dihingia-etal2021}. These facts strongly suggest the presence of BP disc-wind in our simulations.
Although our analysis gives clear indications of two kinds of disc-winds in the accretion flows, it is difficult to quantify the contribution of each mechanism exactly. However, we see that, far from the black hole, the dominant mechanism is BP. By observing the super-Alfv\'enic and sub-Alfv\'enic launching regions only, we will be able to roughly estimate the dominant mechanisms and their dependencies on the input parameters.

In the left panels of Fig.~\ref{fig:disc-wind}a-\ref{fig:disc-wind}d, we study the variation of the signature of $B_{\rm tor}$ and BP disc-wind for different accretion rates. When accretion rates increase, the radiative cooling rates are larger, and the flow becomes much cooler. This reduces the plasma-$\beta$ values in the disc and the disc-surface as compared to low accretion rate cases. As a result, the turbulent nature of the disc-wind becomes less. Thus, we see that the area of $B_{\rm tor}$ dominated disc wind decreases with the increase of the accretion rate (see the bright region of the panels in Fig.~\ref{fig:disc-wind}a-\ref{fig:disc-wind}d). On the other hand, due to the cooler flow close to the equatorial plane (thin-disc), the poloidal component of the magnetic field can be sustained better. As a result, the area of BP disc wind increases with the increase in the input accretion rate (see the dark region of the panels in Fig.~\ref{fig:disc-wind}a-\ref{fig:disc-wind}d). Note that the dependency of the accretion rate is very small on the nature of the magnetised disc wind.

In the right panels of Fig.~\ref{fig:disc-wind}e-\ref{fig:disc-wind}h, we study the role of magnetic field strength on the signature of $B_{\rm tor}$ and BP disc wind. In a previous study, \cite{Dihingia-etal2021} reported that the stronger magnetic fields are more susceptible to BP disc wind, whereas they are less susceptible to $B_{\rm tor}$ disc wind. In our simulations, we also observe that the area of $B_{\rm tor}$ dominated disc-wind decreases with the increase in magnetic field strength (lowering $\beta_0$, see the bright white region). On the other hand, the area of BP disc wind increases with the increase in magnetic field strength (lowering $\beta_0$, see the dark region). Thus, with the inclusion of radiative cooling and heating, we do not find any qualitative alteration in the results reported by \cite{Dihingia-etal2021}. It indicates that magnetised disc wind properties do not depend on radiative cooling and heating.

In all the panels of Fig.~\ref{fig:disc-wind}, we observe some asymmetric structure along the disc mid-plane. Such asymmetry in the disc mid-plane is more prominent in cases with higher magnetic field strengths. For the case of the lowest magnetic field strength ($\beta_0=0.5$), the Alfv\'enic Mach number profile is completely symmetric. Similar properties can be seen in the density structure as well in Fig.~\ref{fig:density}. Earlier, it has been seen that for stronger magnetic field cases, growing instabilities such as magnetic Rayleigh-Taylor instabilities (MRTI) are one of the possible mechanisms for driving the turbulence \citep{Igumenshchev2008, Avara-etal2016, Marshall-etal2018}. Such turbulence close to the black hole may create asymmetric disc wind, which results in asymmetric net pressure across the mid-plane. This could be one of the reasons for the development of the asymmetric structure in the disc mid-plane.

\section{Radiative properties}
In this section, we investigate the radiative properties of the thin disc and the impacts of the input accretion rate and magnetic field strength on them. To do that, we calculate light curves at different emitted frequencies using general-relativistic radiative transfer (GRRT) code \texttt{BHOSS} \citep{Younsi-etal2012,Younsi-etal2020,Younsi2023}. \texttt{BHOSS} solve the covariant radiative transfer equation, which is given by,
\begin{align}
    \frac{d{\cal I}}{d\tau_v}=-{\cal I} + \frac{\zeta}{\chi},
\end{align}
where Lorentz-invariant specific intensity ${\cal I}$ is related to specific intensity ${\cal I}_{\nu}$ at frequency $\nu$ by ${\cal I}={\cal I}_\nu/\nu^3$. The Lorentz-invariant emissivity $(\zeta)$ and absorptivity $(\chi)$ are determined by the emission ($j_\nu$) and absorption coefficients $(\alpha_\nu)$ at frequency $\nu$ such that $\zeta=j_{0,\nu}/\nu^2$ and $\chi=\alpha_{0,\nu}\nu$, where a value of $0$ denotes quantities as observed in the plasma's local rest frame. Using these definitions, we write two partial differential equations for optical depth ($\tau_\nu$) and specific intensity $(\cal I)$ and solve them following by \cite{Younsi-etal2012}. The emission coefficients ($j_{0,\nu}$) are calculated from our GRMHD simulation data and input to the \texttt{BHOSS} for GRRT calculations. Our goal is to understand emissions coming from the inner part of the accretion flows around a stellar-mass black hole in different energy ranges. Accordingly, for our calculation, we consider the emission from bremsstrahlung, thermal synchrotron, and black body radiation following \citep{Rybicki-Lightman1979, Leung-etal2011, Page-Thorne1974}, respectively. For this analysis, emission from black body radiation is not considered self-consistently. In reality, the instantaneous values of the accretion rate of the thin disc vary with time (see Fig.~\ref{fig:acc-mag}). Accordingly, the black body temperature should also vary with time. However, for simplicity, we consider the average accretion rate (target accretion rate) to calculate the black-body temperature throughout.

To calculate the emission, we set the mass of the target sources to the same as the GRMHD simulations, $M_{\rm BH}=10M_\odot$, and set the target accretion rate to be the given $\dot{m}$. In this study, we do not target any specific astrophysical source. Accordingly, we arbitrarily set the distance and inclination angle of the source to be $1,000$ parsec and $i=60$ degree, respectively, with a resolution of $320\times320$ pixels of images with $\pm 20\,r_{g}$ widths. Note that since in our GRMHD simulations, we calculate the electron temperature self-consistently, we do not need to specify any electron temperature prescription. To avoid the emission from the highly magnetised polar regions that affect density floor treatment, we set a cutoff value of magnetisation, $\sigma_{\rm cut}=1$ \citep[e.g.,][]{EHTCV}, beyond which we neglect all the emission. Additionally, we compute emission only from close to the black hole ($r \le 20\,r_g$), and the emission from the outside is neglected. 
Since we calculate the emission only close to the black hole, therefore, we calculate the emission within the simulation time $t=13000-15000\,t_g$.

\begin{figure}
    \centering
    \includegraphics[scale=0.35]{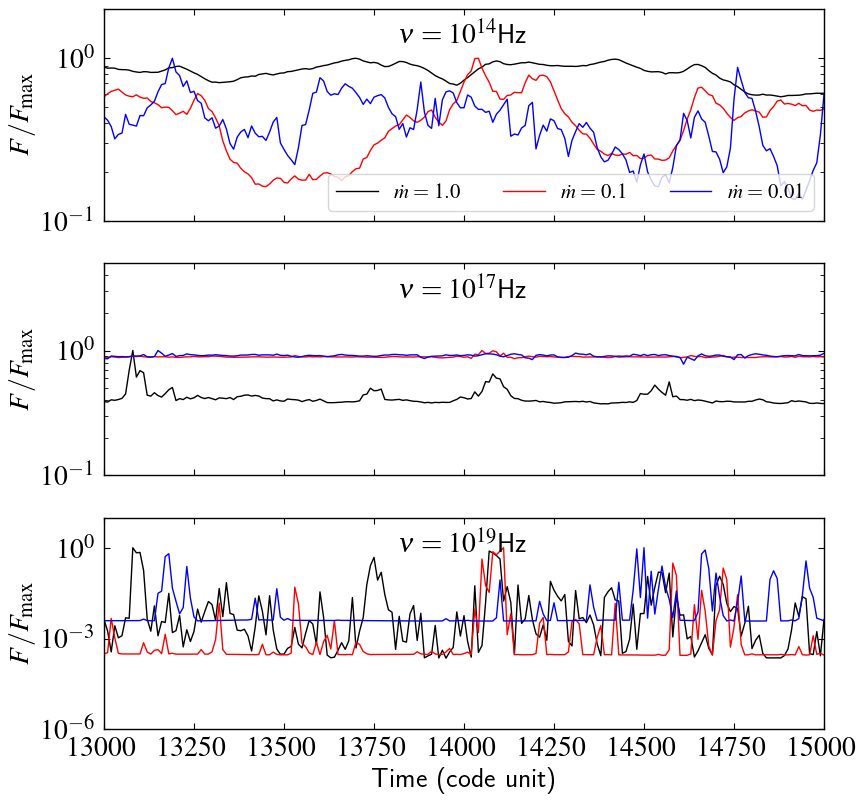}
    \caption{Plot of light-curves at frequencies $\nu=10^{14}, 10^{17}$ and $10^{19}$Hz for different values of the 
    accretion rates $(\dot{m})$ marked on the figure. 
    }
    \label{fig:LC-mdot}
\end{figure}

We show the light curves at frequencies $\nu=10^{14}, 10^{17}$ and $10^{19}$Hz for different accretion rates in Fig.~\ref{fig:LC-mdot}. For better comparison, we express the light curves in terms of normalized flux $F(Jy)/F_{\rm max}(Jy)$. Black, blue, and red lines correspond to light curves at accretion rates $\dot{m}=0.1, 0.01$, and $0.001$, respectively. These frequencies are chosen in such a way that for the frequency $\nu=10^{14}$Hz, the synchrotron process dominates. For $\nu=10^{17}$~Hz, the black body component dominates, and for $\nu=10^{19}$Hz, either synchrotron or bremsstrahlung dominates. At frequency $\nu=10^{14}$Hz, the average value of the photon flux for accretion rates $\dot{m}=0.1, 0.01$, and $0.001$ are $\langle F(\nu=10^{14}) \rangle=0.016, 8.8\times10^{-4},$ and $7.9\times10^{-6}$ Jy, respectively. The temporal properties show a very small variability pattern at the frequency. At frequency $\nu=10^{17}$Hz, the average value of the flux for accretion rate $\dot{m}=0.1, 0.01$, and $0.001$ are $\langle F(\nu=10^{17}) \rangle =0.028, 4.6\times10^{-4},$ and $4.4\times10^{-6}$ Jy, respectively. The temporal properties of the light curves at this frequency for different accretion rates do not show any variability, as these fluxes mostly come from the black body radiation, which comes from the steady equatorial part of the flow. At frequency $\nu=10^{19}$Hz, we observe quite a similar feature as compared to frequency $\nu=10^{17}$Hz, with a more prominent variability pattern. The average value of the photon flux at this frequency for accretion rates $\dot{m}=0.1, 0.01$, and $0.001$ are $\langle F(\nu=10^{19}) \rangle =9.4\times10^{-6}, 5.4\times10^{-8},$ and $8.1\times10^{-11}$ Jy, respectively. However, the light curves at the lower end are smoothed out for accretion rates $\dot{m}= 0.01$ and $0.001$, essentially suggesting that the emission from the Bremsstrahlung process is higher than that of the Synchrotron process at this frequency. We also note that the emission from Bremsstrahlung does not vary significantly with time, as the Bremsstrahlung radiation is emitted only from the high-density thin-disc region, which remains dynamically steady throughout the simulation time. On the other hand, high-energy synchrotron radiation is emitted mostly from the disc-wind region, which is turbulent in nature. As a result, synchrotron-dominated light curves are much more variable (see the next section for details).

\begin{figure}
    \centering
    \includegraphics[scale=0.35]{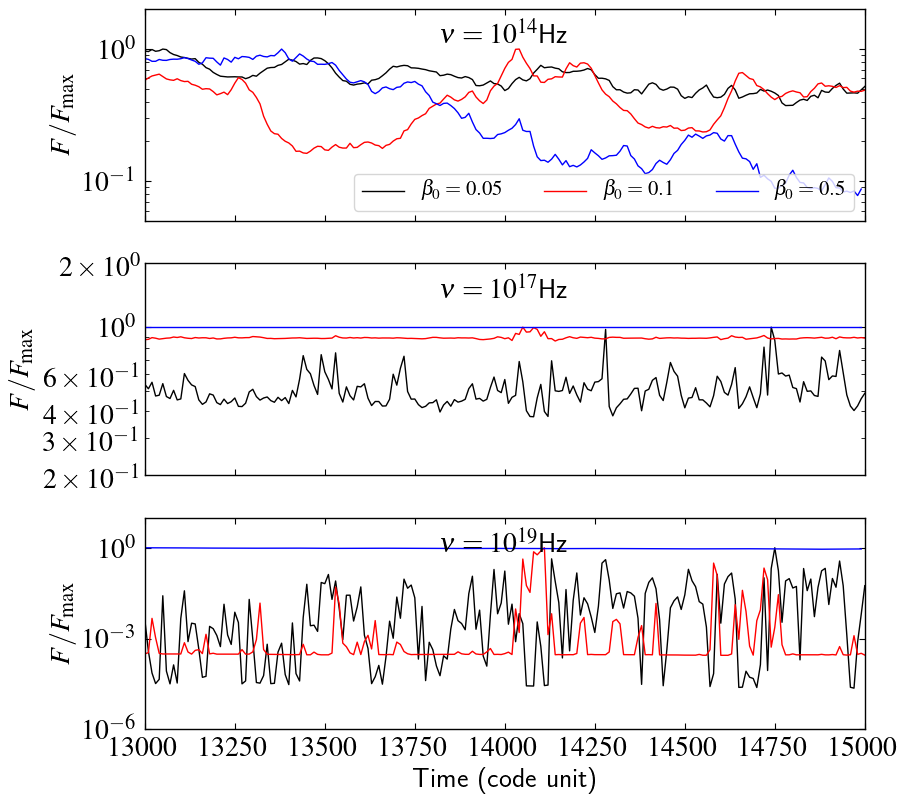}
    \caption{Plot of light-curves at frequencies $\nu=10^{14}, 10^{17}$ and $10^{19}$Hz for different values of the 
    initial magnetic field strength $(\beta_0)$ marked on the figure.}
    \label{fig:LC-beta}
\end{figure}

Figure~\ref{fig:LC-beta} shows the comparison of light-curves for different initial plasma-$\beta$ ($\beta_0$) parameters at frequencies ($\nu=10^{14}, 10^{17},$ and $10^{19}$Hz. We see that for weak magnetic field limits ($\beta_0=0.5$), the photon flux at all frequencies is very small as compared to moderate ($\beta_0=0.1$) and strong magnetic field limits ($\beta_0=0.05$). At frequency $\nu=10^{17}$Hz, the light curves for $\beta_0=0.5$ and $0.1$ are quite similar and steady. As for this magnetic field strength, the synchrotron emission is always weaker than that of the black body component. However, for $\beta_0=0.05$, we observe a variable light curve with an average flux greater than that of the weak magnetic field cases. This suggests stronger synchrotron emission compared to black body radiation. In the case of $\nu=10^{19}$Hz and $\beta_0=0.5$, the photon flux is very small $\langle F(\nu=10^{19}) \rangle \sim 10^{-12}$ and does not vary at all with time, which is dominated only by the Bremsstrahlung radiations. However, in other cases, the flux and the variability pattern increase with the increase in magnetic field strength (lowering $\beta_0$).


\begin{figure}
    \centering
    \includegraphics[scale=0.42]{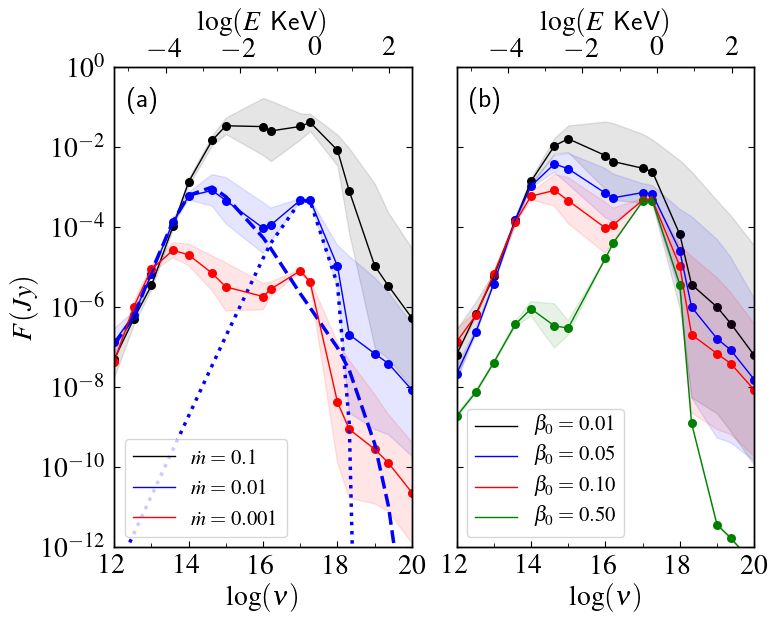}
    \caption{Plot of the time-averaged spectrum for different accretion rate (left, $\dot{m})$ and initial plasma-$\beta$ 
    parameters (right, $\beta_0)$. The shaded area shows the spread of the photon flux throughout the simulation time. The blue dashed and dotted lines in panel (a) correspond to the contributions from the synchrotron and the black body process.}
    \label{fig:sed}
\end{figure}

\subsection{Average spectra}

To understand the impacts of the accretion rates and the initial plasma-$\beta$ parameters on the radiative properties throughout the energy range, in Fig.~\ref{fig:sed}, we show the time-averaged spectra for different accretion rates (left, $\dot{m})$ and initial plasma-$\beta$ parameters (right, $\beta_0)$. For the left figure, the initial plasma-$\beta$ is fixed at $\beta_0=0.1$ and for the right figure, the accretion rate is fixed at $\dot{m}=0.01$. In the figure, a shaded region indicates the spread of photon flux throughout the temporal domain ($t=13000-15000\,t_g$). In the top `x-axis', for the benefit of readers, we show the photon energy $E=h\nu$ in terms of $\mathrm{keV}$. To understand the contribution of the individual emission processes, we overlay the contribution of the synchrotron and the black body in dashed and dotted lines for $\dot{m}=0.01$ case (blue colour) in Fig.~\ref{fig:sed}a. The spectra from the individual processes are calculated by ray tracing them separately. Thus, the very low energy part ($E<\sim0.1$keV) is dominated by the synchrotron emission, whereas the intermediate range $E\sim 0.1-10$keV is dominated by the black body component. The very high energy $E>\sim10$keV is dominated by the Bremsstrahlung process. All the spectra show the characteristic synchrotron peak around the $\nu\sim 10^{15}-10^{16}$Hz frequency range. In the frequency range $\nu\sim 10^{16}-10^{18}$Hz ($\sim 0.1-10$keV), we observe the characteristic peak for the black body radiation. In the higher frequency range $\nu>10^{18}$Hz, we observe the characteristic signature of the Bremsstrahlung radiation, which is more prominent for lower accretion rates and weaker magnetic field strength. 

With the increase in the accretion rate, the magnetic field strength increases due to the scaling relationship. As a result, the synchrotron emission increases with the higher accretion rate. On the other hand, Bremsstrahlung emission also increases with the increase in density. Subsequently, in the figure, we see that the photon flux increases with the increase in the accretion rate. Also, the peak frequency of the radiation shifts towards the higher frequency range. Similarly, with the accretion rate, the peak due to the black body radiation also sifts towards the higher energy range with increases in flux in orders of magnitude. For higher accretion rates, the high-energy tail at high frequency is also dominated by synchrotron radiation due to the presence of a stronger magnetic field. For lower accretion rates, the Bremsstrahlung radiation dominates the high-energy tail.

With the decrease of initial plasma-$\beta$ parameters $(\beta_0)$, the magnetic field strength increases monotonically (see Fig.~\ref{fig:mag-energy}). Subsequently, the total magnetic flux increases with the decrease of $\beta_0$, and the peak synchrotron frequency shifts towards the higher frequency range. However, as the accretion rate is fixed, the black body component does not change at all with the increase in magnetic field strength. Nonetheless, in the case of lower magnetic field strength, dominating the emission process is black body radiation. Accordingly, we observe that the second peak for $\beta_0=0.5$ is much higher than the first synchrotron peak (four orders of magnitudes). With the increase in magnetic field strength, the flux from the synchrotron emission becomes comparable with the black body emission, and eventually it overpowers the black body emission for the very strong magnetic field case ($\beta_0=0.01$). This suggests that emission peaks in the soft X-ray region from a thin-accretion disc may be from synchrotron radiation if the magnetic field is strong enough. Although, while varying the $\beta_0$, we fixed $\dot{m}$ for the simulation models, we observe a significant increase in the emission from the Bremsstrahlung process. This fact essentially suggests that for weaker magnetic field limits (higher $\beta_0$) the electrons remain cold in the flow. Whereas, for a stronger magnetic field limit (lower $\beta_0$) electrons are heated up by the turbulent heating mechanism. For this reason, the emission from the Bremsstrahlung process increases significantly with an increase in magnetic field strength. 

Throughout this study, we observe that the variability of the light curves around $\nu\sim10^{17}$Hz is minimal. This is an artifact of considering the average accretion rate to calculate the black body temperature. With self-consistent calculations, we expect some variability in this range. We plan to do such calculations in the future.

\subsection{Coronal temperature}
Throughout our calculation, we have not considered any contribution to the emission from the inverse Compton process. However, the contribution from the inverse Compton process is inevitable. In Fig.~\ref{fig:temp_e}, we observe that surrounding the funnel, there is a region with a very hot electron temperature. The soft electrons from the black body component can interact with the hot electrons and can energise via the inverse Compton process in this region. Also, the emission due to the synchrotron process may re-energise via the self-Compton process. These processes are very important to understand the hard x-ray part of the emission spectra (e.g., \cite{Steiner-etal2009, Titarchuk-etal2014, Poutanen-etal2018}, etc.). 

\begin{figure}
    \centering
    \includegraphics[scale=0.28]{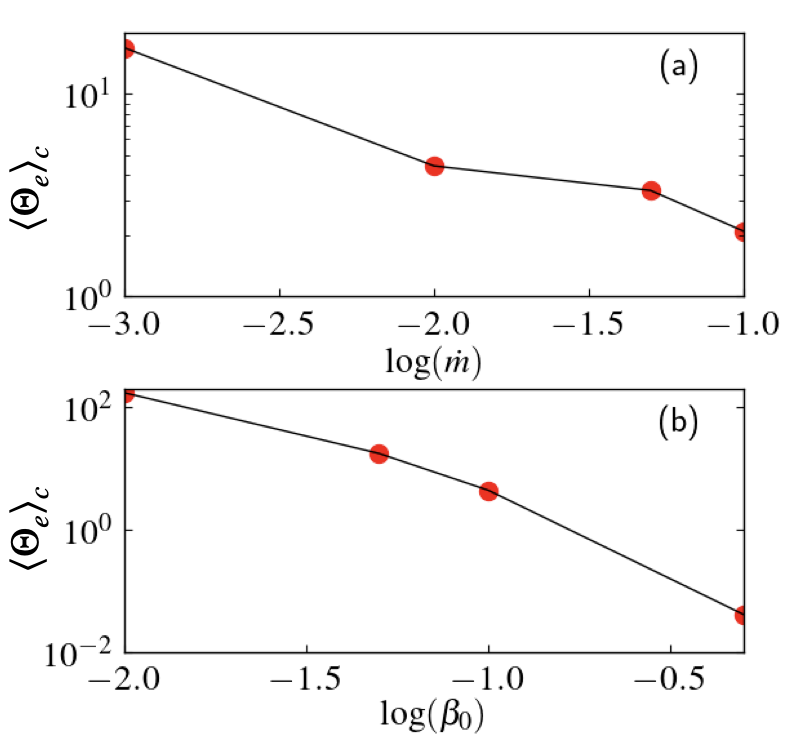}
    \caption{Average coronal temperature is shown as a function of (a) input accretion rate $(\dot{m})$ and the (b) 
    input magnetic field strength ($\beta_0$).}
    \label{fig:corona-temp}
\end{figure}

The GRRT \texttt{code BHOSS} is currently not equipped to handle the Compton process; therefore, in this work, we calculate the average electron temperature of the corona region $(\langle \Theta_e \rangle_c)$ following,

\begin{align}
    \langle \Theta_e \rangle_c = \frac{\sum_k \rho^k \Theta_e^k }{\sum_k \rho^k},
\end{align}
where the summation $(k)$ is carried over all the simulation grids with $-hu_t>1$. To avoid inconsistency due to the different sizes of the grids in SMR, the summation is performed after converting SMR grids to uniform grids. The Compton efficiency is directly related to the coronal temperature. The higher value of coronal temperature results in a stronger Compton hump, and the lower corona temperature results in a weaker Compton hump in the emission spectra \citep[e.g.,][etc.]{Titarchuk1994,Vurm-Poutanen2009,Zdziarski-etal2020}. The plots for time-averaged value $\langle \Theta_e \rangle_c$ for different values of accretion rates ($\dot{m}$) and input plasma-$\beta$ ($\beta_0$) are shown in Fig.~\ref{fig:corona-temp}. The figures essentially suggest that with the increase in accretion rate, the coronal temperature decreases. This is due to the fact that the radiative efficiency increases with the accretion rate and therefore it could cool down the coronal temperature.
The value of $\langle \Theta_e \rangle_c$ drops from $16.93$ to $2.1$ when the accretion rate is increased from $\dot{m}=0.001$ to $\dot{m}=0.1$. On the other hand, with the decrease in magnetic field strength (increase in $\beta_0$), the average value of $\langle \Theta_e \rangle_c$ decreases. This suggests that the turbulent magnetic heating decreases with the decrease of magnetic field strengths. Accordingly,
the value of $\langle \Theta_e \rangle_c$ drops from $173.1$ to $0.042$ when $\beta_0$ is decreased from $\beta_0=0.1$ to $\beta_0=0.5$.

In summary, we found that both parameters $\dot{m}$ and $\beta_0$ impact the radiative properties of the thin-disc significantly. From the above discussion, it is evident that an accretion disc with a higher accretion rate and lower magnetic field strength corresponds to a soft spectral state with dominating spectral energy $E>10$keV. With the decrease in accretion rate or the increase in magnetic field strength, the coronal temperature increases, resulting in a soft-intermediate to hard-intermediate spectral state depending on the accretion rate and magnetic field strength.

\section{Conclusions and discussion}
In this work, we have investigated radiatively cooled thin accretion discs around black holes in the presence of turbulent heating, Coulomb interaction, Bremsstrahlung, synchrotron, Comptonisation of synchrotron photons, and black body radiation. With this, we examined the role of accretion rate and magnetic field strength on the dynamics as well as the radiative properties of the accretion flow. We also studied the properties of the disc-winds with radiative cooling. Below, we summarised our concluding remarks point-wise. 

\begin{itemize}
    \item We find that all of our simulation models remain within the SANE accretion flow limit except for the very strong magnetic field case (model \texttt{G}), which becomes MAD with temporal evolution. The SANE models with different input accretion rates show different accretion rates through the event horizon. However, they show similar magnetic flux accumulations around the black hole. On the contrary, the magnetic flux accumulated around the black hole is different for the SANE models with different magnetic field strengths, although the accretion rate through the event horizon is similar. 
    
    \item The input accretion rate and the magnetic field strength both play crucial roles in deciding the properties of the accretion flow. For a higher accretion rate, the electrons are cooler throughout the domain. In comparison with the low accretion rate case, the electrons in the equatorial plane also become cooler in the high accretion rate case. With the increase in magnetic field strength, the matter in the thin disc puffed up. However, in the case of a very strong magnetic field, the matter in the thin disc evaporates faster due to the onset of strong disc winds. Also, with the increase in magnetic field strength, the electrons in the accretion flow become hotter, particularly in the corona region. 
    
    \item We observe the presence of the $\beta\ll1$ region above the disc surface, where the ratio is $T_e/T_i\gg1$ . As a result, the usual $R-\beta$ prescription for describing electron temperature in terms of flow/ion temperature (e.g., \cite{Moscibrodzka-etal2016,Dihingia-etal2023}) cannot provide the trend of electron temperature in the case of a thin disc. Such descriptions are useful only for geometrically thick accretion flow cases (e.g., see \cite{Dihingia-etal2023}).

    \item We find the ratio of temperatures $T_e/T_i$ significantly depends on the electrons heating prescriptions. For turbulent heating, the electrons in the funnel region are much hotter compared to the ions. However, in the case of reconnection heating, the electrons in the funnel region mostly remain cooler than the ions.

    \item We find ample evidence for the active BP disc wind on the thin disc surface. The active area of the BP disc wind increases with the increase in input accretion rate as well as the magnetic field strength. The mass flux rate due to the disc wind also increases with the input accretion rate and the magnetic field strength. 

    \item By studying the light curves coming from the accretion flow for different input accretion rates and magnetic field strengths, we found that the intensity of the light curves increases both with the accretion rate as well as the magnetic field strength. The light curves at higher frequencies show much more variable features as compared to the light curves at lower frequencies. In the case of Bremsstrahlung dominated over synchrotron (high frequency, low accretion rate, low magnetic field strength), the light curves do not show variability features.

    \item The time-averaged spectra also show the importance of the accretion rate and the magnetic field strength in deciding the radiative properties of the accretion flow. The area under the spectra, or the total luminosity, increases with both the accretion rate as well as the magnetic field strength. For higher accretion rates and lower magnetic field cases, the black body component dominates the peak of the emission spectra. For lower accretion rates and higher magnetic field cases, the synchrotron dominates the peak of the emission spectra.
    
    \item To understand the contribution of the inverse-Compton process in the emission spectra, we calculate the average coronal temperature $(\langle \Theta_e \rangle_c)$. We find that $\langle \Theta_e \rangle_c$ is anti-correlated with the accretion rate, whereas it is correlated with the strength of the magnetic field. 
\end{itemize}

Based on our study, we find that the accretion rate and the magnetic field strength both play a vital role in deciding the spectral states of an accretion flow. We summarise our understanding of accretion flow in terms of accretion rate limits, magnetic field strength limits, possible spectral states, and characteristics of the outflow in table \ref{tab-02}. The table shows a possible link between the accretion rates and the magnetic field strength with the disc winds. Many studies have speculated such a bell-shaped behaviour of the accretion rate profile during an outburst \citep[e.g.,][]{Esin-etal1997,Ferreira-etal2006,Kylafis-Belloni2015,Kumar-Yuan2022}. Also, observations of MAXI J1820+070 show polarised emission in the rising hard state, whereas they show unpolarised emission during the soft and decaying hard states. These facts hint that magnetic field configuration is possibly structured and strong during the rising phase, and the same is weak and unstructured in the decaying phase. Such observations can only be comprehended if the magnetic field contribution varies over the timeline of the outburst. At the end, we want to mention that our simulations are performed with axisymmetric consideration. It is well known that asymmetric MRI can not keep dynamo going for a long time (e.g., Cowling's anti-dynamo theory \cite{Cowling1933}). Accordingly, a full 3D simulation is better for studying the long-term behaviour of accretion flows. To make our study consistent, we only simulate up to $t=30,000\,t_g$, however with a mean-field dynamo approach, we could accomplish 3D dynamo effects with less numerical cost \citep[e.g.,][]{Parker1955, Krause1980, Sadowski-etal2015}. We plan to do such studies in the future. 
Finally, in this study, the impacts of radiation in the accretion flow are considered in terms of radiative cooling, which is a straightforward and computationally efficient \citep[e.g.,][]{Fragile-Meier2009,Dibi-etal2012, Yoon-etal2020}. However, despite being far more computationally expensive, an M1 closure \citep{Sadowski-etal2017, Chael-etal2018, Chael-etal2019} or a Monte-Carlo scheme \citep{Ryan-etal2017,Ryan-etal2018,Dexter-etal2021} treatments are more suitable for examining the fluid dynamics coupling with the radiation field. We plan to implement such complicated schemes in our future studies. Additionally, we would like to mention that due to the short simulation time ($30000\,t_g~147.6\times M_{\rm BH}/10M_\odot$ milliseconds) and the absence of Compton scattering process, we do not study the detailed x-ray variability features (e.g., QPOs) in this study. Although they are crucial aspects for deciphering physics, BH-XRBs in the outburst state can be studied separately in the future with the appropriate framework.
\begin{table}
\centering
  \begin{tabular}{| c |c | c | c |}
    \hline
    Accretion rate & Field Strength & States & Disc-winds\\ 
    \hline
    Low ($\lesssim 0.01\dot{M}_{\rm Edd}$)  &  High  & High Hard &  Structured  (BP) \\
    High ($\gtrsim \dot{M}_{\rm Edd}$) &  Low   & Soft &  Turbulent $(B_{\rm tor})$ \\
    Low ($\lesssim 0.01\dot{M}_{\rm Edd}$) &  Very low   & Low Hard &  Turbulent $(B_{\rm tor})$ \\
    \hline
  \end{tabular}
\caption{Summary of parameters ranges and corresponding spectral states and disc-wind characteristics.}
\label{tab-02}
\end{table}

\appendix
\section{MRI Quality factor}

\begin{figure}
    \centering
    \includegraphics[scale=0.48]{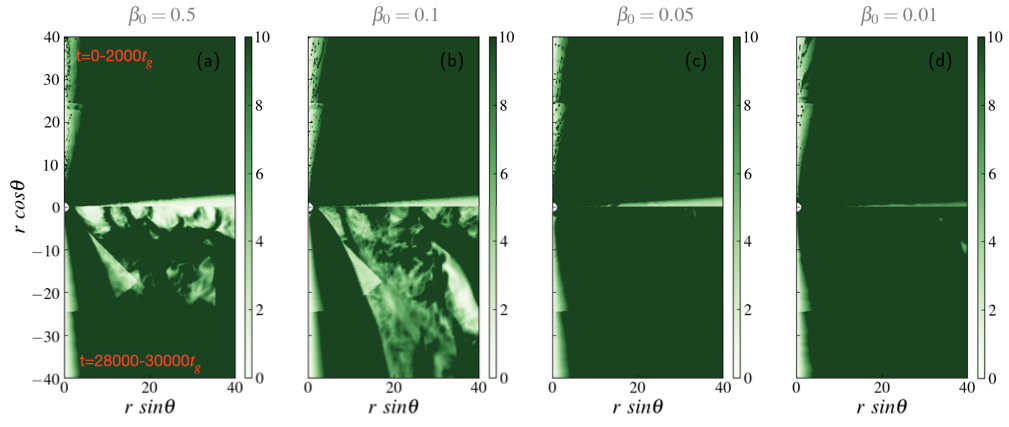}
    \caption{Time averaged distribution of $Q_\theta$ for different simulation models with different magnetic field strengths. The upper and lower half of the figure corresponds to the time-averaged distribution over $t=0-2000\,t_g$ and $t=28000-30000\,t_g$, respectively. See text for more details.}
    \label{fig:qtheta}
\end{figure}

To ascertain the status of MRI resolution for a given numerical model, we define quality factor in terms of the wavelength of the fastest growing MRI mode $\lambda_\theta$ along the $\theta$ direction as $Q_\theta = \lambda_\theta/\Delta x_\theta$ (see \cite{Takahashi2008,Siegel-etal2013,Porth-etal2019,Nathanail-etal2020}, for details). Here, $\lambda_\theta$ is given by
\begin{align}
\lambda_\theta = \frac{2\pi}{\sqrt{(\rho h + b^2)\Omega}}b^\mu e_\mu^{(\theta)},
\end{align}
and the grid resolution $\Delta x_\theta = \Delta x^\mu e_\mu^{(\theta)}$. These quantities are calculated at tetrad basis of the fluid frame $e_\mu^{\hat{(\alpha)}}$. This MRI mode is typically resolved with $Q_\theta \gtrsim 6$  (see \cite{Sano-etal2004}). In Fig. \ref{fig:qtheta}, we show the value of $Q_\theta$ for different simulation models with different magnetic field strengths (fixed accretion rate), i.e., model \texttt{C, E, F}, and \texttt{G} in panels (a)-(d), respectively. The values of $\beta_0$ are marked on the top of each panel. The upper half of the figure corresponds to the time-averaged distribution over $t=0-2000\,t_g$, while the lower half corresponds to the time-averaged distribution over $t=28000-30000\,t_g$. The figure suggests that the MRI resolved well ($Q_\theta\gtrsim10$) in most of the simulation domains, irrespective of the simulation time. However, for model \texttt{C} and \texttt{E}, the MRI is slightly under-resolved ($Q_\theta\sim2-4$) in the high-density equatorial region.  

\acknowledgments
We appreciate the thoroughness of the thoughtful comments provided by the anonymous reviewer, which have improved the manuscript. This research is supported by The National Key R\&D Program of China (grant No. 2023YFE0101200), the National Natural Science Foundation of China (grant No. 12273022), and
the Shanghai Municipality orientation program of basic research for international scientists (grant No. 22JC1410600).  CMF is supported by the DFG research grant ``Jet physics on horizon scales and beyond" (Grant No.  FR 4069/2-1).
ZY acknowledges support from a United Kingdom Research \& Innovation (UKRI) Stephen Hawking Fellowship.
The simulations were performed on the TDLI Astro cluster, Pi2.0, and Siyuan Mark-I at Shanghai Jiao Tong University.
This work has made use of NASA's Astrophysics Data System (ADS). 

{\bf Data Availability:}
The data underlying this article will be shared on reasonable request to the corresponding author.




\end{document}